\documentclass[preprintnumbers,nofootinbib,noshowpacs,eqsecnum,prd,superscriptaddress,letterpaper]{revtex4-2}
\pdfoutput=1
\usepackage{graphicx}
\usepackage{amsmath,amssymb}
\usepackage{url}
\usepackage{multirow}
\usepackage{hyperref,color}
\usepackage[normalem]{ulem}
\usepackage{slashed}
\usepackage{rotating}
\usepackage{afterpage}
\usepackage{ifthen}
\usepackage{mciteplus}
\usepackage{multirow}
\usepackage{color}
\usepackage{pifont}

\def\lsim{\raise0.3ex\hbox{$\;<$\kern-0.75em\raise-1.1ex\hbox{$\sim\;$}}}
\def\gsim{\raise0.3ex\hbox{$\;>$\kern-0.75em\raise-1.1ex\hbox{$\sim\;$}}}

\newcommand{\eh}{\hat{e}}
\newcommand{\sh}{\hat{s}}
\newcommand{\ch}{\hat{c}}
\newcommand{\vh}{\hat{v}}

\newcommand {\qbw}     {{Q}_{BW}}
\newcommand {\qpone}   {{Q}_{\Phi,1}}

\newcommand {\qww}     {{Q}_{WW}}
\newcommand {\qbb}     {{Q}_{BB}}

\begin{document}

\preprint{YITP-SB-2026-08}
\title{USMEFT as a tool for discovery of universal new physics at high luminosity LHC}
\author{Tyler Corbett}
\email{corbett.t.s@gmail.com}
\affiliation{Faculty of Physics, University of Vienna, Boltzmanngasse 5, A-1090 Wien, Austria}
\author{Jay Desai}
\email{jay.desai@stonybrook.edu}
\affiliation{C.N. Yang Institute for Theoretical Physics,
  Stony Brook University, Stony Brook New York 11794-3849, USA}
\author{O.\ J.\ P.\ \'Eboli}
\email{eboli@if.usp.br}
\affiliation{Instituto de F\'{\i}sica, 
Universidade de S\~ao Paulo, S\~ao Paulo -- SP  05508-090, Brazil.}

\author{M.~C.~Gonzalez-Garcia}
\email{concha.gonzalez-garcia@stonybrook.edu}
\affiliation{C.N. Yang Institute for Theoretical Physics,
  Stony Brook University, Stony Brook New York 11794-3849, USA}
\affiliation{Departament de Fis\'{\i}ca Qu\`antica i
  Astrof\'{\i}sica and Institut de Ciencies del Cosmos, Universitat de
  Barcelona, Diagonal 647, E-08028 Barcelona, Spain}
\affiliation{Instituci\'o Catalana de Recerca i Estudis
  Avan\c{c}ats (ICREA), Pg. Lluis Companys 23, 08010 Barcelona,
  Spain.}
\author{Matheus Martines}
\email{matheus.martines.silva@usp.br}
\affiliation{Instituto de F\'{\i}sica, 
Universidade de S\~ao Paulo, S\~ao Paulo -- SP  05508-090, Brazil.}
%
\begin{abstract}
  We analyze the potential of the Universal SMEFT as a tool to study
  universal new physics in the neutral and charged Drell-Yan processes
  at the high-luminosity LHC.  In order to do so, we generated
  pseudo-data containing the contributions of beyond the standard
  model vector bosons to the dilepton production.  Our results show
  that fits with the minimal theoretical bias can be used not only to
  unveil the existence of new physics but also to accurately extract
  its properties. Moreover, the results are rather stable with respect
  to the order of truncation of the effective field theory.
\end{abstract}

\maketitle

\section{Introduction}

Up to now there has been no direct evidence of new physics (NP) at the
CERN Large Hadron Collider (LHC) despite the accumulated
statistics. One possibility is that the states associated to the new
dynamics are too heavy to produce a resonance signature at the LHC but
they are not too far from the kinematic reach.  In this scenario it is
standard to look for NP signals in the tail of kinematic
distributions, with effective field theory (EFT) being the optimum
model-independent theoretical tool to perform such analysis. \smallskip

In this context, dilepton production, {\em i.e.}  the Drell-Yan (DY)
process~\cite{Drell:1970wh}, is one of the most precisely measured
observables at the LHC.  It can proceed either via neutral current
(NC) or charged current (CC)
\[
  p p  \to \ell^+ \ell^-    \;\;\;\hbox{ and }\;\;\; p p \to \ell^\pm \nu_\ell
\]
with $\ell =e,\mu$.  Involving only leptons in the final state,
provides a clean environment for both experimental studies and
theoretical predictions. Thus it constitutes one of the most promising
windows for beyond the standard model (BSM) searches. In fact the LHC
experimental collaborations have taken advantage of these features to
perform precision standard model (SM)
tests~\cite{ATLAS:2016gic,CMS:2014jea, Walia:2018fon,
  CMS:2024ony,ATLAS:2025hhn} and searches for new
resonances~\cite{ATLAS:2019erb, CMS:2021ctt, ATLAS:2019lsy,
  CMS:2022krd}. In the near future the high-luminosity run of the LHC
(HL-LHC) is expected to accumulate 3000 fb$^{-1}$ at 14 TeV, enlarging
the parameter space that can be probed. \smallskip

As mentioned above, a particularly suitable approach to parametrize
deviations in kinematic distributions in a model-independent fashion
is to employ the EFT framework. Under the minimal assumption that the
scalar particle observed in 2012~\cite{Aad:2012tfa, ATLAS:2017jag} is
part of an electroweak doublet, the $SU(2)_L \otimes U(1)_Y$ symmetry
can be linearly realized and the resulting EFT is the so-called
Standard Model EFT (SMEFT). In this framework, deviations from the SM
predictions are parametrized as higher order operators
\begin{equation}
  {\cal L}_{\rm eff}  = {\cal L}_{\rm SM} + \sum_{n>4,j}
  \frac{f_{n,j}}{\Lambda^{n-4}} {\cal O}_{n,j} \;, 
\end{equation}
where $\Lambda$ is a characteristic high energy scale, ${\cal O}_{n,j}$ are
higher dimensional operators, and $f_{n,j}$ are their respective Wilson
coefficients (WC). \smallskip

In the bottom-up approach to BSM searches in this context, WCs are treated as
being independent in the fit to the considered dataset. Open questions
in this framework include: Can such an EFT fit with arbitrary WC provide a clear
signature of NP? If so, do the extracted values of the WC's
capture the properties of the BSM? Do the answers to these two questions
depend on the order of the truncation of the EFT series? \smallskip

These questions are central to the EFT methodology and as such
they have been approached in the existing literature. In this respect,
the study of correlations in the SMEFT series associated with
specific UV completions as well as the challenge of identifying the specific UV
completion in SMEFT fits has been previously addressed 
(see for example ~\cite{Dawson:2020oco,Dawson:2021xei,Ellis:2020unq,Dawson:2024ozw, Adhikary:2025gdh, Dawson:2022cmu, Ellis:2023zim,Corbett:2024evt,Brivio:2021alv,terHoeve:2023pvs,Greljo:2023adz,Greljo:2023bdy,Hirsch:2025qya}
for an incomplete list).
There is also an increasing number of  studies of the impact of dimension-eight
operators in EFT analyses either stressing the existence of new
signal topologies due to dimension-eight operators~\cite{Liu:2016idz, 
Degrande:2013kka, Alioli:2020kez} or investigating the
changes generated by the inclusion of these operators to fits
performed at dimension six, in Drell-Yan
~\cite{Alte:2018xgc, Boughezal:2021tih, Boughezal:2022nof,
  Kim:2022amu, Allwicher:2022gkm, Allwicher:2022mcg,
  Allwicher:2024mzw, Corbett:2025oqk}, production of electroweak gauge
bosons~\cite{Liu:2018pkg, Degrande:2023iob, Corbett:2023qtg,
  Martin:2023tvi, Gillies:2024mqp, Goldberg:2024eot}, Higgs
physics~\cite{Hays:2018zze, Assi:2024zap,
  Corbett:2021cil,Martin:2021cvs,Corbett:2021jox, Hays:2020scx,
  Flores-Hernandez:2026sac, Bera:2026ild}, and electroweak precision
observables~\cite{Hays:2020scx, Corbett:2021eux}.
Somehow none of these studies directly addressed the question of the
discovery of NP in fits 
with minimal theoretical priors, such us SMEFT fits, which is the
focus of our work. \smallskip

Thus it is with these questions in focus, that we assess in this work the HL-LHC
potential to search for NP employing the SMEFT framework. We consider
two simplified scenarios in which we extend the SM with the addition of
a new $U(1)$ gauge field in one case, and additional $SU(2)$ vector
bosons in the other.  As is well-known, beyond
${\cal O}(1/\Lambda^2)$, the SMEFT in its most general form possesses
an untreatable number of operators which precludes a definitive answer
to the third question above. The answer becomes accessible, however, in
the context of Universal BSM scenarios, that is, when the NP either
dominantly couples to the SM bosons and if it couples to fermions,
does so via SM currents.  Therefore, we will consider two scenarios
where the new gauge bosons couple to SM currents, also referred as
mirror-hypercharge (or mirror-$U(1)_Y$) and mirror-$SU(2)_L$ models.
\smallskip

Our strategy to quantitatively answer the above questions proceeds as
follows. First we simulate pseudo-data containing the SM and BSM
predictions for the DY observables in the context of the two models
for a variety of model parameters (masses and couplings). In addition
we also take into account the electroweak precision observables (EWPO)
in order to better constrain the parameter space under consideration.
Subsequently we perform a fit to the DY pseudo-data and EWPO for each
model and each set of model parameters using USMEFT and obtained the
best fit and allowed range of all the operator Wilson coefficients
included in the fit.  In this framework, the signal of NP is, then,
associated with the determination of some non-vanishing Wilson
coefficients within a particular confidence level. The ability to
identify the specific NP is addressed by comparing the extracted
ranges of coefficients from the fit with the predictions from matching
the specific model to the USMEFT operators.  Furthermore, in order to
address the dependence of the answers on the order of the EFT
truncation, we perform the analyses including all operators at orders
$1/\Lambda^2$ and $1/\Lambda^4$.  \smallskip

The presentation of our work is organized as follows. In Section.~\ref{sec:effoper}
we summarize  the  USMEFT operators relevant for the analyses of DY
data up to order $1/\Lambda^4$, complemented by
Appendix~\ref{app:couplings}. The analyzed BSM scenarios are
introduced in Sec.~\ref{sec:bsm} with further details in
the Appendix~\ref{app:fullmatching}. The specifics of the
framework and assumptions in our our analyses are contained in
Sec.~\ref{sec:framework}.  Our results are presented in detail 
in Sec~\ref{sec:results} and we briefly summarize  our
findings in Sec.~\ref{sec:summary}.

\section{Operator basis}
\label{sec:effoper}

Within the USMEFT predictions for observables at order
${\cal O} (1/\Lambda^4)$ require evaluating the SM contributions, the
interference between the $1/\Lambda^2$ amplitude (${\cal M}^{(6)}$)
with the SM amplitude, the square of the dimension-six amplitude, as
well as the interference of the $1/\Lambda^4$ amplitude with the SM,
which we represent as:
\begin{equation}
  | M_{\rm SM}|^2 + {\cal M}_{\rm SM}^\star {\cal M}^{(6)}  + | {\cal
    M}^{(6)}|^2 + {\cal M}_{\rm SM}^\star {\cal M}^{(6,2)}+ {\cal M}_{\rm SM}^\star {\cal M}^{(8)}
  \;.
\label{eq:ampli}  
\end{equation}
${\cal M}^{(8)}$ includes amplitudes with one dimension-eight operator
coefficient while ${\cal M}^{(6,2)}$ includes the contribution of the
insertion of two dimension-six Wilson coefficients in the
amplitude. \smallskip

In order to perform our analyses we must choose a basis of independent
operators for the USMEFT. Under the assumption of universality an
independent basis can be constructed of {\sl bosonic} operators, that
is operators formed exclusively of bosonic fields.  As is widely
known, operators connected by the use of the classical equations of
motion (EOM) of the SM fields lead to the same $S$--matrix
elements~\cite{Politzer:1980me, Georgi:1991ch,
  Arzt:1993gz,Simma:1993ky}. Thus, it is possible to trade bosonic
operators with higher derivatives for operators involving fermions in
the form of combinations of SM currents herein called {\sl fermionic
  operators}. We refer to the resulting basis of this exchanged as the
{\sl rotated basis}. In fact in order to perform the Monte Carlo event
simulations it is more convenient to work with the operators in the
rotated basis ~\cite{Corbett:2023qtg,Brivio:2014pfa}.  \smallskip

The complete list of universal dimension-six operators is presented in
Ref.~\cite{Wells:2015uba}. Assuming that the Yukawa couplings are
negligible as well as requiring $C$ and $P$ conservation one can
identify six independent bosonic operators contributing to the weak
boson propagators at order $1/\Lambda^2$, hence contributing to EWPO
and/or Drell-Yan processes. Two of these bosonic operators can be
rotated into fermionic operators by the EOM to form the relevant
rotated set of dimension-six operators listed in
Table~\ref{tab:uniopd6}, where $H$ stands for the SM Higgs doublet and
${W}_{\mu\nu}^I $ and ${B}_{\mu\nu}$ are the $SU(2)_L$ and $U(1)_Y$
field strength tensors respectively.  $\tau^I$ are the Pauli matrices.
As mentioned above, the fermionic operators involve the SM fermion
currents,
\begin{equation}
  J_B^\mu = g' \displaystyle\sum_{f \in \{q,l,u,d,e\}} \sum_a Y_f
  \bar{f}_a\gamma^\mu f_a \, 
  \;\;\; \hbox{ and }\;\;\;
  J^{I\mu}_{W} = \frac{g}{2} \displaystyle\sum_{f \in \{q,l\}}
  \sum_a \bar{f}_a\gamma^\mu \tau^I f_a\;,\label{eq:sm-cur}
\end{equation}
with $Y_f$ standing for the fermion $f$ hypercharge, $q$ and $l$ are
the quark and lepton doublets and $u$, $d$ and $e$ represent the
fermion singlets. The sum over $a$ is over generations while the
$SU(2)_L$ and $U(1)_Y$ gauge couplings are $g$ and $g^\prime$
respectively. \smallskip

\begin{table}
\begin{tabular}{|l||l|l|}
  \hline
 \multicolumn{2}{|c|}{Operator Basis}&
\text{Coefficient} 
 \\\hline
$\qpone$&  $(D_\mu H^\dagger H)(H^\dagger D^\mu H)$ &
$c_{\Phi,1}$\\
$\qww$ &$H^{\dagger} {W^I}_{\mu \nu} {W}^{I,\mu \nu} H$ &
$c_{WW}$\\ 
$\qbb$ &  $H^{\dagger} {B}_{\mu \nu} {B}^{\mu \nu} H$
&$c_{BB}$\\
$\qbw$ &$H^\dagger {B}_{\mu\nu}\tau^I {W}^{I, \mu\nu} H$ &
${c_{BW}}$\\\hline
$Q_{2JW}$ &  $J^I_{W\mu} J^{I\mu}_W$ &
$c_{2JW}$\\
$Q_{2JB}$ &  $J_{B\mu} J^\mu_B$ & $c_{2JB}$\\
\hline
\end{tabular}
\caption{C and P conserving dimension-six operators that contribute
  to DY and EWPO in the rotated basis for universal theories and their
  respective Wilson coefficients. }
\label{tab:uniopd6}
\end{table}

The full basis of dimension-eight operators for universal theories was
presented in Ref.~\cite{Corbett:2024yoy} and contains 175 operators of
which 86 bosonic operators with higher derivatives can be rotated into
fermionic operators by the EOM as shown in
Ref.~\cite{Corbett:2024yoy}. From those we identify 11 independent
bosonic operators contributing to the weak boson propagators of which
8 can be rotated by the EOM into combinations involving 10 fermionic
operators.  We list in Table~\ref{tab:uniopd8} those independent
dimension-eight operators relevant for EWPD and Drell-Yan. The EOM
imply two relations connecting the coefficients of the following
operators in the {\sl rotated basis}
\begin{equation}
c^{(1)}_{\psi^2 H^2 D^3}=-\frac{g'}{2} c^{(2)}_{\psi^4 D^2}  \;, \;\;\;\;
c^{(2)}_{\psi^2 H^2 D^3}=-\frac{g}{2} c^{(3)}_{\psi^4 D^2}\;.
\label{eq:linkuniversal}
\end{equation}
%

\begin{table}
\begin{tabular}{|l||l|l|}
\hline
\multicolumn{2}{|c|}{{Operator Basis}} & {Coeff}  \\\hline
$Q^{(2)}_{H^6} $ & $ (H^\dagger H) (H^\dagger
\tau^I H) (D_\mu H)^\dagger \tau^I D^\mu H
$ & $ c^{(2)}_{H^6}$\\[+0.1cm]
$Q_{WBH^4}^{(1)} $ & $  (H^\dag H) (H^\dag \tau^I H) W^I_{\mu\nu} B^{\mu\nu}
$ & $ c^{(1)}_{BWH^4} $ \\[+0.1cm]
$Q_{W^2H^4}^{(3)}
$ & $ (H^\dag \tau^I H) (H^\dag \tau^J H)W^I_{\mu\nu} W^{J\mu\nu}  
$ & $ c^{(3)}_{W^2H^4} $ \\[+0.1cm]
\hline
  $Q^{(2)}_{\psi^4 D^2} $ & $
  D^\alpha  J_B^\mu D_\alpha J_{B\mu}
$ & $c^{(2)}_{\psi^4 D^2}$
  \\[+0.1cm]
$Q^{(1)}_{\psi^2 H^2D^3} $ & $
\begin{array}{l}    i\,(D^\mu J_B^{\nu}+D^\nu J_B^{\mu}) \\
  \times (D_{(\mu}D_{\nu)} H^\dagger H-H^\dagger D_{(\mu}D_{\nu)} H)
\end{array}
  $ & $\begin{array}{l}   
    c^{(1)}_{\psi^2H^2 D^3}
    \\
  \end{array}$\\[+0.4cm]
$   Q^{(3)}_{\psi^4 D^2} $ & $  D^\alpha J_W^{I\mu} D_\alpha J_{W\mu}^I
 $ & $ c^{(3)}_{\psi^4 D^2}$\\[+0.1cm]
 $ Q^{(2)}_{\psi^2 H^2D^3} $ & $
  \begin{array}{l}
      i\,(D^\mu J_W^{I\nu}+D^\nu J_W^{I\mu}) \\\times (D_{(\mu}D_{\nu)}
  H^\dagger\tau^I H-H^\dagger\tau^I D_{(\mu}D_{\nu)}H)
\end{array}
   $ & $ \begin{array}{l}
     c^{(2)}_{\psi^2H^2 D^3}
    \\
  \end{array}$\\[+0.4cm]
$ Q^{(1)}_{\psi^2 H^4 D} $ & $
i\,J_B^\mu (H^\dagger\overleftrightarrow{D}_{\mu} H) (H^\dagger H)
$ & $c^{(1)}_{\psi^2 H^4 D} $\\ [+0.1cm]            
$ Q^{(2)}_{\psi^2 H^4 D} $ & $   \begin{array}{r}i\,J_W^{I\mu}\left[ (H^\dagger\overleftrightarrow{D}^I_{\mu} H) (H^\dagger H)\right.\\\left.+(H^\dagger\overleftrightarrow{D}_\mu H) (H^\dagger \tau^I H)\right]\end{array}          
$ & $c^{(2)}_{\psi^2 H^4 D}$ \\[+0.1cm]
$ Q^{(4)}_{\psi^2 H^4 D} $ & $
\epsilon^{IJK}\,J_W^{I\mu}(H^\dagger\tau^J_{\mu} H) D_\mu(H^\dagger \tau^K H)
$ & $c^{(4)}_{\psi^2 H^4 D}$\\         [+0.1cm]    
$ Q^{(4)}_{\psi^4 H^2} $ & $
 J_B^{\mu}J_{B\mu}   (H^\dagger H)
$ & $c^{(4)}_{\psi^4 H^2} $\\                 [+0.1cm]
$ Q^{(5)}_{\psi^4 H^2} $ & $
 J_W^{I\mu}J_{W\mu}^I  (H^\dagger H)
$ & $c^{(5)}_{\psi^4 H^2} $\\                 [+0.1cm]
$ Q^{(7)}_{\psi^4 H^2} $ & $
J_W^{I\mu}J_{B\mu} (H^\dagger\tau^I H)
$ & $c^{(7)}_{\psi^4 H^2} $ \\                 
\hline
\end{tabular}
\caption{C and P conserving USMEFT dimension-eight operators that
  contribute to DY and EWPO in the rotated basis and their respective
  Wilson coefficients.}
\label{tab:uniopd8}
\end{table}

As discussed in Ref.~\cite{Corbett:2025oqk} the EWPO and DY analyses
can not constrain all 17 Wilson coefficients that enter the analyses
at order ${\cal O} (1/\Lambda^4)$.  However, in
Ref.~\cite{Corbett:2025oqk} we showed that after carefully accounting
for both the direct contributions and the indirect effects induced by
the finite renormalization of the SM parameters~\cite{Corbett:2023qtg}
up to ${\cal O}(\Lambda^{-4})$, one can construct five effective
combinations of Wilson coefficients
\begin{eqnarray}
 \overline{c}_{BW}
&\equiv&
 c_{BW} \left(1-{ \frac{\vh^2}{\Lambda^2}} (c_{WW}+c_{BB})\right) + 
 \frac{1}{2}\left[ c^{(1)}_{WBH^4} 
     +\frac{\eh}{2\sh}
     c^{(1)}_{\psi^2 H^4D} +\frac{\eh}{\ch}
     c^{(2)}_{\psi^2 H^4D}\right]\frac{\vh^2}{\Lambda^2} \;,
\label{eq:cbwb}\\
 \overline{c}_{\Phi,1}
& \equiv& {c}_{\Phi,1}
+\left[c_{H^6}^{(2)}-\frac{\eh}{\ch}c^{(1)}_{\psi^2 H^4D}
    -\frac{\eh}{\sh}
    \big(c^{(2)}_{\psi^2 H^4D}-c^{(4)}_{\psi^2 H^4D}\big)
    \right]\frac{\vh^2}{\Lambda^2}\;,
\label{eq:cp1b}\\
\overline{c}^{(3)}_{W^2H^4}
& \equiv &c^{(3)}_{W^2H^4}+\frac{\eh}{2\sh}\left(
c^{(2)}_{\psi^2 H^4D}-c^{(4)}_{\psi^2 H^4D}\right)
     \label{eq:c3whb}
      \overline{c}_{2JW} \;,
\\
 \overline{c}_{2JW} &\equiv& c_{2JW}\left(1{-2\frac{\vh^2}{\Lambda^2} c_{WW}}\right)
  + { c^{(5)}_{\psi^4 H^2} \frac{\vh^2}{2 \Lambda^2}}
  \equiv -\frac{2\sh^2}{\eh^2}\overline{\Delta}_{4F} \;,
\label{eq:c2jwb}
  \\  \overline{c}_{2JB}&\equiv& c_{2JB}\left(1-2\frac{\vh^2}{\Lambda^2}c_{BB}\right)
  + { c^{(4)}_{\psi^4 H^2}\frac{\vh^2}{2\Lambda^2}} \;.
\label{eq:c2jbb}  
\end{eqnarray}
which together with the coefficients of the dimension-eight operators
$c^{(7)}_{\psi^4 H^2}$, $c^{(2)}_{\psi^4 D^2}$, and
$c^{(3)}_{\psi^4 D^2}$, can be independently determined by studying
the invariant mass distribution of the lepton pairs produced in DY
processes. \smallskip

For the sake of completeness, we summarize the basic elements and the
final expressions for the fermion couplings to the electroweak gauge
bosons and the four-fermion amplitudes relevant for the EWPO and DY
analyses expressed in terms of these eight effective coefficients in
Appendix~\ref{app:couplings}. In particular, notice that in
Eq.~\eqref{eq:c2jwb} we have introduced $\overline{\Delta}_{4F}$ which
is an effective parameter defined in the literature containing the
contribution of fermionic operators to the muon decay width (see
Eq.~\eqref{eq:vt}).  \smallskip

\begin{table}
\begin{center}
  \resizebox{\textwidth}{!}{%
\begin{tabular}{|c||c| c| c|| c| c| c|}
  \hline
  & \multicolumn{3}{c||}{Mirror-$U(1)_Y$} &
  \multicolumn{3}{c||}{Mirror-$SU(2)_L$} \\\hline
Effective Coefficients
& dim-6 & dim-6 squared&  dim-8 &  dim-6 & dim-6 squared&  dim-8 \\
\hline
$\overline{c}_{BW}$
& $\displaystyle-\frac{\hat e^2 \beta^2}{4\hat s\hat c}$
& ${ \displaystyle -\frac{\hat e^4 \hat v^2 \beta^4}{8\hat s_{4W}M_1^2}}$
& $ \displaystyle -\frac{\hat e^4 \hat v^2 \beta^2}{16M_1^2\hat c^3\hat s} (1+\beta^2)$
& $\displaystyle -\frac{\hat e^2\beta^2}{4\hat s\hat c}$
& ${\displaystyle -\frac{\hat e^4 \hat v^2 \beta^4}{64\hat s^3\hat c M_2^2}\left(\frac{1}{\hat c_{2W}}-3\right) }$
& $\displaystyle -\frac{\hat e^4 \hat v^2 \beta^2}{16\hat s^3 \hat c M_2^2} $
 \rule{0pt}{18pt}\\
$\overline{c}_{\Phi,1}$
& $\displaystyle\frac{\hat e^2 \beta^2}{2\hat c^2}$
& ${-\frac{\hat e^4\hat v^2\beta^4}{4\hat c^4}}$
& $0$
& $0$
& $0$
& $0$
 \rule{0pt}{18pt}\\
$\overline{c}_{2JB}$
& $\displaystyle-\frac{\beta^2}{2}$
& $\displaystyle-\frac{\hat e^2\hat v^2\beta^4}{4\hat c^2 M_1^2}$
& $\displaystyle+\frac{\hat e^2\hat v^2\beta^4}{8\hat c^2 M_1^2}$
& $0$
& $0$
& $0$
 \rule{0pt}{18pt}\\
$\overline{c}_{2JW}\equiv-\frac{2\hat s^2}{e^2} \overline{\Delta}_{4F}$
& $0$
& $0$
& $0$
& $\displaystyle-\frac{\beta^2}{2}$
& $\displaystyle -\frac{\hat e^2 v^2 \beta^4}{4\hat s^2 M_2^2}$
& $\displaystyle +\frac{\hat e^2 v^2 \beta^4}{8\hat s^2 M_2^2}$
 \rule{0pt}{18pt}\\
$\overline{c}^{(3)}_{W^2H^4}$
& -
& -
& $\displaystyle-\frac{\hat e^4 \beta^2}{16\hat s^2\hat c^2} (1+\beta^2)$
& - 
& - 
& $0$
\\
$c^{(7)}_{\psi^4 H^2}$
& -
& -
& $0$
& -
& -
& $0$
\\
$c^{(2)}_{\psi^4 D^2}$
& -
& -
& $\displaystyle -\frac{\beta^2}{2}$
& -
& -
& $0$
\\
$c^{(3)}_{\psi^4 D^2}$
& -
& -
& $0$
& -
& -
& $\displaystyle -\frac{\beta^2}{2}$
\\\hline
\end{tabular}
}
\end{center}
  \label{tab:WC}
  \caption{Low-energy Wilson coefficients in the rotated basis
    generated by integrating out the gauge bosons of the $mirror-U(1)_Y$ and
    mirror-$SU(2)_L$ models that contribute to DY and EWPO.}
\end{table}

%
\section{BSM scenarios}
\label{sec:bsm}

In order to study the potential of the HL-LHC to unveil signals of
universal NP we study two specific models. The first is the
``mirror-$U(1)_Y$ model'' where a copy of the SM $U(1)_Y$ field is
assumed, the second is a mirror-$SU(2)_L$ model where we instead study
a copy of the SM $SU(2)_L$ field. In both cases the SM fields have the
same charges under the new $U(1)$ ($SU(2)$) gauge field as to the SM
gauge field for $U(1)_Y$ ($SU(2)_L$). \smallskip

In this section we discuss how the matching up to dimension-eight
operators is achieved for both of these models. In order to do so we
establish our conventions by writing out the relevant Lagrangians
coupling the NP to the SM degrees of freedom. This allows us to, in
later sections, compare the ultraviolet (UV) behavior of the model
where the NP degrees of freedom are allowed to propagate with that of
the infrared (IR) theory or EFT where the new degrees of freedom have
been integrated out.

\subsection{ Mirror-$U(1)_Y$ Model}

We begin by discussing the mirror-$U(1)_Y$ model. In this case we have
a duplicate of the SM $U(1)_Y$ field we write as $X^\mu$. For a
generic SM field $F$, the covariant derivative of the field in the UV
is given by:
\begin{align}
\bar D_\mu F =& \left(\partial_\mu -ig' Y_F B_\mu-ig' \beta Y_F X_\mu+\cdots\right)F\, ,\\
=&D_\mu F-ig'\beta Y_F X_\mu F\, .
\end{align}
In this expression the ellipsis represents the coupling to the other
SM gauge bosons. We have defined the covariant derivative in the UV
theory to be $\bar D$ and that involving only SM gauge bosons to be
$D$.  As our goal is to integrate out the $X$ field, it is more
convenient to factor the $X$-dependence out of the covariant
derivative.  Notice that, rather than introducing a separate gauge
coupling, we measure the strength of the coupling of $X$ in units of
$g'$ with the coupling scaling $\beta$, {\em i.e.} $g_X=g'
\beta$. \smallskip

For the fermionic sector the explicit $X$ field dependence is given by:
\begin{align}
i\bar\psi\slashed{\bar D}\psi =& i\bar\psi\slashed{D}\psi +\beta X_\mu J_B^\mu\, ,
\end{align}
where $J_B$ is one of the SM fermionic currents appearing in
Eq.~\eqref{eq:sm-cur}. The fact that the new $X$ field only couples to
the SM fermions through this current is precisely the statement that
this UV theory is ``Universal''. The covariant derivative on the right
does not have a bar as it involves only the SM gauge bosons.
Performing the same for the Higgs kinetic term yields:
\begin{align}
(\bar D_\mu H)^\dagger (\bar D^\mu H)=&(D_\mu H)^\dagger (D^\mu H)+ig'\beta Y_H(H^\dagger \overleftrightarrow D_\mu H)X^\mu+g'^2\beta^2 Y_H^2(H^\dagger H)X_\mu X^\mu\, \\
\equiv&(D_\mu H)^\dagger (D^\mu H)+g'\beta Y_H \mathcal H_\mu X^\mu+g'^2\beta^2 Y_H^2(H^\dagger H)X_\mu X^\mu\, .
\end{align}
The last term, proportional to $X^2$, is irrelevant for dimension-six matching and therefore commonly omitted \cite{Madigan:2024cky,Hammou:2023heg}. In our case  we must retain this term as it is relevant for matching at dimension-eight. Above we have implicitly defined the Higgs current we use for simplifying our expressions below:
\begin{align}
\mathcal H^\mu =i(H^\dagger \overleftrightarrow D_\mu H)\, .
\end{align}
Taking into account the above discussion our full UV Lagrangian is given by:
\begin{align}
\mathcal L=\mathcal L_{\rm SM}-\frac{1}{4}X_{\mu\nu}X^{\mu\nu}+\frac{1}{2}M_1^2X_\mu X^\mu+\beta X_\mu J_B^\mu+g'\beta Y_HX_\mu\mathcal H^\mu+g'^2\beta^2(H^\dagger H)X_\mu X^\mu\, .
\end{align}
We do not include a kinetic mixing term, $X_{\mu\nu} B^{\mu\nu}$, as
in \cite{Dawson:2024ozw}. This is because, in the case of the mirror
model where the charges of the SM fields under the new $U(1)$ are the
same as under the SM $U(1)$, we can remove the mixing term by a field
redefinition \cite{Corbett:2024evt}.  The mass, $M_1$, denotes
specifically the explicit mass appearing in the mirror-$U(1)_Y$
Lagrangian above. The physical masses in the UV model, used for the
full theory simulations, are calculated below the electroweak
  symmetry breaking (EWSB) scale after diagonalizing mass mixing
effects. We present our calculation for the resulting heavy and light
neutral gauge boson masses up to order $1/M_1^2$, while the full
analytic expression is omitted due to its length:
\begin{align}
\overline M_1^2=&M_1^2\left(1+\frac{g'^2 v^2}{4M_1^2}\beta^2+\cdots\right)\, ,\\
\overline m_Z^2=&\frac{(g'^2+g^2)v^2}{4}\left(1-\frac{g'^2v^2}{4M_1^2}\beta^2+\cdots\right)\, .
\end{align}

We perform the tree-level matching up to dimension-eight using the
Matchete package \cite{Fuentes-Martin:2022jrf}.  The results of the
matching were checked by hand by solving the classical EOM for the $X$
field as described in \cite{Henning:2014wua}.  The dimension-six
effective operators that we obtain are
\begin{equation}
    \mathcal{L}_{\textrm {SM}} + \mathcal{L}_6 \rightarrow
    \mathcal{L}_{\textrm {SM}} -\frac{\beta^2}{2M_1^2}Q_{2JB} +
    \frac{g'^2}{M_1^2}\beta^2 Y_H^2 (Q_{\Phi,2} - 2Q_{\Phi,1}) -
    i\frac{g'^2}{M_1^2} \beta^2Y_H Y_\psi(H^\dagger
    \overleftrightarrow{D}_\mu H) (\bar{\psi} \gamma_\mu \psi) \,.
\end{equation}
The last operator needs to be traded for operators in the
\textsl{rotated} basis. To consistently include effects up to
${\cal O} (1/\Lambda^4)$, it is insufficient to use the EOM, and we
need to perform field redefinitions \cite{Criado:2018sdb}.  So, we
make the following field redefinition:
\begin{equation}
B_\mu\to B_\mu +\frac{g'\beta^2}{M_1^2}Y_H\mathcal H^\mu\, ,
\end{equation}
This redefinition was made by hand and cross checked by manipulating
the Matchete output in Mathematica. We take into account linear
redefinitions of the dimension-six lagrangian and quadratic terms in
the field redefinitions of the renormalizable lagrangian to capture
the dimension-eight terms.  We find for the dimension-six Lagrangian:
\begin{equation}
\mathcal L_6=-\frac{\beta^2}{2M_1^2}Q_{2JB}+\frac{g'^2}{M_1^2}\beta^2Y_H^2\left(2Q_{\Phi,1}-Q_{\Phi,2}\right)-\frac{g'}{M_1^2}\beta^2Y_H\left[2Q_B+\frac{g'}{2}Q_{BB}+\frac{g}{2}Q_{BW}\right]\, .
\end{equation}
The operators, $Q_B = i (D_\mu H)^\dagger B^{\mu\nu} (D_\nu H)$ and
$Q_{\Phi,2} = \tfrac{1}{2} \partial_\mu (H^\dagger H)\partial^\mu
(H^\dagger H)$, do not contribute to EWPO or the Drell-Yan process at
tree level. \smallskip

The resulting dimension-eight Lagrangian arising from the field
redefinitions as well as tree-level matching was then reduced by
making extensive use of Matchete's internal integration-by-parts
routines in order to align the dimension-eight operators with the
purely bosonic basis presented in \cite{Corbett:2024yoy}. These
results were then rewritten in the Murphy basis \cite{Murphy:2020rsh}
following the relations presented in \cite{Corbett:2024yoy}. The part
of the dimension-eight Lagrangian relevant to our analysis of
Drell-Yan in the rotated basis is then given by:
\begin{align}
  \mathcal{L}_{8} =&  -\frac{g^{2}g^{\prime \,
    2}\beta^{2}}{8M_1^4}(1+\beta^{2})\,Q^{(2)}_{H^{6}}   -\frac{gg^{\prime \, 2}\beta^{2}}{8M_1^4}(1+\beta^{2})\,Q^{(2)}_{\psi^{2}H^{4}D}   +\frac{g^\prime\beta^{2}}{4M_1^4}\,Q^{(1)}_{\psi^{2}H^{2}D^{3}}\nonumber\\
    &  -\frac{\beta^{2}}{2M_1^4}\,Q^{(2)}_{\psi^{4}D^{2}}
+\frac{g^{\prime \, 2}\beta^{4}}{4M_1^4}\,Q^{(4)}_{\psi^{4}H^{2}}+\cdots\;.
\label{eq:L8U1reduced}
\end{align}
The ellipses denotes 19 additional dimension-eight operators that do
not contribute to the Drell-Yan observables considered in this
work. The full result of the matching can be found in
Eq.~\eqref{eq:L8U1} in App.~\ref{app:fullmatching}. Rewriting these
results in terms of our input parameter scheme and relating them to
the barred coefficients as defined in
Eqs.~\eqref{eq:cbwb}--\eqref{eq:c2jbb} yields the relations found in
Table~\ref{tab:WC}.

\subsection{Mirror-$SU(2)_L$ Model}

Next we consider new heavy gauge bosons, $X^a_\mu$, corresponding to a
mirror $SU(2)$ symmetry. The SM fermionic and Higgs doublets are also
doublets under this mirror symmetry and we obtain the following UV
Lagrangian:
\begin{equation}
    \mathcal{L} = \mathcal{L}_{SM} - \frac{1}{4} (X^a_{\mu\nu})^2 +
    \frac{1}{2}M_2^2(X^a_{\mu})^2 + \beta X^a_\mu J_W^{a \mu}  + \frac{g}{2}\beta
    X^{a,\mu} \mathcal{H}^a_\mu+
    \frac{g^2}{4}\beta^2 (H^\dagger H)X^{a,\mu}X^a_{\mu} \;,
\end{equation}
where, again, $\beta$ measures the strength of the mirror-$SU(2)$
gauge coupling relative to the coupling $g$ of the SM.  In this case
the mass of the new field is denoted $M_2$ in order to distinguish it
from that of the mirror-$U(1)_Y$ model. Again, in the full UV model
this mass is shifted by mixing effects. To order $1/M_2^2$ the masses
of the resulting charged and neutral vectors are given by:
\begin{align}
\left(\overline M_2^{\pm,0}\right)^2=&M_2^2\left(1+\frac{g^2v^2}{4M_2^2}\beta^2+\cdots\right)\, ,\\
\overline m_W^2=&\frac{g^2v^2}{4}\left(1-\frac{g^2v^2}{4M_2^2}\beta^2+\cdots\right)\, ,\\
\overline m_Z^2=&\frac{(g'^2+g^2)v^2}{4}\left(1-\frac{g^2v^2}{4M_2^2}\beta^2+\cdots\right)\, .
\end{align}
The Higgs current appearing above is defined as:
\begin{align}
    \mathcal{H}^a_\mu = i(H^\dagger \overleftrightarrow{D}^a_\mu H) =
    i \left[ H^\dagger \tau^a
D_\mu H - (D_\mu H)^\dagger \tau^a H\right]\;,
\end{align}
while the fermionic current appears in Eq.~\eqref{eq:sm-cur}.\smallskip

We integrate out the heavy gauge bosons using Matchete as described in
the case of the mirror-$U(1)$ model. In order to rewrite the
dimension-six operator basis we first make the $W$ field redefinition,
\begin{align}
W_\mu^a\to W_\mu^a+\frac{g}{2M_2^2}\beta^2\mathcal H_\mu^a\, ,
\end{align}
then subsequently redefine the Higgs doublet as,
\begin{align}
H\to H-\frac{g^2}{4M_2^2}\beta^2H|H|^2\, . \label{eq:Higgsredef}
\end{align}
The resulting dimension-six Lagrangian is given by:
\begin{equation}
  \mathcal{L}_{6} = \frac{g^2\beta^2}{M_2^2}\left(\lambda +
    \frac{\beta^2g^2\mu^2}{16M_2^2}\right)Q_{\Phi^6} 
  - \frac{\beta^2}{2M_2^2}Q_{2JW} -
  \frac{3g^2\beta^2}{4M_2^2}Q_{\Phi,2}   - \frac{g
    \beta^2}{M_2^2} Q_W  - \frac{g^2}{4M_2^2}\beta^2 Q_{WW}
  - \frac{g'g}{4M_2^2}\beta^2 Q_{BW} \;,
\end{equation}
where $\lambda$ is the quartic coupling of the Higgs potential.  The
operators $Q_W = i (D_\mu H)^\dagger \tau^I W^{I\,\mu\nu} (D_\nu H)$,
$Q_{\Phi,2}$, and $Q_{\Phi^6} = (H^\dagger H)^3$ do not contribute to
either EWPO or the Drell-Yan process at tree level. The Higgs field
redefinition in Eq.~\eqref{eq:Higgsredef} also results in a shifted
quartic Higgs operator, however our analysis is insensitive to this
shift.\smallskip

After rewriting the dimension eight operators in terms of the rotated
basis we find the operators relevant to our analysis are:
\begin{equation}
\begin{aligned}
\mathcal{L}_8 =
&-\frac{ g^2 g^{\prime\,2}\, \beta^2}{2M_2^4}\,(1+\beta^2)\, Q^{(2)}_{H^6}
+\frac{ g^3 g^\prime\, \beta^2}{8M_2^4}\, Q^{(1)}_{W B H^4}\\
&-\frac{ g^2 g^\prime\beta^2}{2M_2^4}\, (1+\beta^2)\, Q^{(1)}_{\psi^2 H^4 D}
+\frac{g^3\, \beta^4}{4M_2^4}\, Q^{(2)}_{\psi^2 H^4 D}
+\frac{g^3\, \beta^4}{4M_2^4}\, Q^{(4)}_{\psi^2 H^4 D}
+\frac{g\, \beta^2}{4M_2^4}\,  Q^{(2)}_{\psi^2 H^2 D^3}
\\
&+\frac{3g^2\, \beta^2}{4M_2^4}\, Q^{(3)}_{\psi^4 H^2}
+\frac{g^2\, \beta^4}{4M_2^4}\, Q^{(5)}_{\psi^4 H^2}
-\frac{\beta^2}{2M_2^4}\, \, Q^{(3)}_{\psi^4 D^2}+\cdots\;.
\end{aligned}
\end{equation}
The full matching results including operators that do not affect our
analyses (denoted by the ellipsis above) can be found in
App.~\ref{app:fullmatching}. The results expressed as barred
coefficients in terms of the input parameters appear in
Tab.~\ref{tab:WC}.

\section{Analysis Framework}
\label{sec:framework}

As a starting step, prior to our HL-LHC studies, we obtain the
parameter regions of the BSM models already excluded at 95\%
confidence level (CL) by the presently available data. Since each of
the BSM scenarios considered in this work generates a single oblique
parameter: $Y$ for the mirror-$U(1)_Y$ model and $W$ for the
mirror-$SU(2)_L$ model, the strongest bounds arise from the high-$p_T$
Drell–Yan tails as emphasized previously in
Ref.~\cite{Farina:2016rws}.  Therefore, we perform an analysis of the
Run 1 and 2 LHC data in Table~\ref{tab:dydata} in the context of these
two models.  We obtain the following limits on the model parameters:
\begin{align}
    \, {\rm mirror}-U(1)_Y:&\, |\beta| \le 0.23\, \dfrac{ M_1}{\text{TeV}} \quad
       [95\%\, \text{CL}]\,, \\[0.5em] \,{\rm mirror}-SU(2)_L:& \, |\beta| \le
       0.11\, \dfrac{ M_2}{\text{TeV}} \quad [95\%\, \text{CL}]\,.
\end{align}

\begin{table}
\begin{tabular}{|c|c|c|c|c|c|c|}
  \hline
  Channel & Distribution & \# bins & ranges  & data set & Int. Lum.
  \\
  \hline
  NC  & $\frac{d^2\sigma}{dm_{\ell\ell}d|y|_{\ell\ell}}$ & 48&
  $116 \text{\;GeV} \leq m_{\ell\ell}\leq 1.5 \text{\;TeV}$
  $0 \leq y_{\ell\ell}\leq 2.4$ 
  &ATLAS 8 TeV  & 20.3 fb$^{-1}$~\cite{ATLAS:2016gic}
  \\
  \hline
  NC &  $\frac{dN_\text{ev}}{dm_{e^+e^-}}$ & 20 & $250 \text{\;GeV} \leq m_{e^+e^-}\leq 5 \text{\;TeV}$ &
  ATLAS 13 TeV & 139 fb$^{-1}$~\cite{ATLAS:2019erb}
\\
  NC &  $\frac{dN_\text{ev}}{dm_{\mu^+\mu^-}}$ & 20 & $250 \text{\;GeV} \leq m_{\mu^+\mu^-}\leq 5 \text{\;TeV}$ &
  ATLAS 13 TeV & 139 fb$^{-1}$~\cite{ATLAS:2019erb}
\\
NC &  $\frac{dN_\text{ev}}{dm_{e^+e^-}}$ & 20 & $300 \text{\;GeV} \leq m_{e^+e^-}\leq 6 \text{\;TeV}$ 
& CMS 13 TeV & 137 fb$^{-1}$~\cite{CMS:2021ctt}
\\
  NC &  $\frac{dN_\text{ev}}{dm_{\mu^+\mu^-}}$ & 20 & $300 \text{\;GeV} \leq m_{\mu^+\mu^-}\leq 7 \text{\;TeV}$ &
 CMS 13 TeV & 137 fb$^{-1}$~\cite{CMS:2021ctt}\\
  \hline
  CC & $\frac{d\sigma}{dm_T}$ & 20 & $200 \text{\;GeV} \leq m_{T,\ell \nu}
  \leq 5 \text{\;TeV}$ & ATLAS 13 TeV & 140 fb$^{-1}$~\cite{ATLAS:2025hhn}                                                      \\  
CC & $\frac{d N}{dm_T}$ & 20 & $440 \text{\;GeV} \leq m_{T,e
                               \nu}\leq 7 \text{\;TeV}$ & CMS 13 TeV &
                                                                      138 
                                                          fb$^{-1}$~\cite{CMS:2022krd} \\
CC & $\frac{d N}{dm_T}$ & 20 & $600 \text{\;GeV} \leq m_{T,\mu\nu}\leq 7
                               \text{\;TeV}$ & CMS 13 TeV & 138 
                                                          fb$^{-1}$~\cite{CMS:2022krd} \\
  \hline
\end{tabular} 
\caption{Neutral- and charged-current Drell-Yan data to obtain the
  present USMEFT bounds.}
\label{tab:dydata}
\end{table}

These current bounds on the models delimit the target region of model
parameters for our HL-LHC forecasts. For those, as sketched in the
introduction, our starting point is the generation of pseudo-data for
the DY processes following the predictions of the BSM models for grid
of values of the model parameters, {\em i.e.} the masses and
couplings, and confront it with the expectations within the USMEFT
including all the relevant operator coefficients. To this end the BSM
model pseudo-data as well as the USMEFT predictions were obtained with
MadGraph5\_aMC@NLO~\cite{Frederix:2018nkq} at leading order in QCD and
QED.  The required UFO files for the two BSM models considered as well
as the USMEFT relevant operators were created with the help of
FeynRules~\cite{Christensen:2008py, Alloul:2013bka}.  Parton shower
and hadronization was performed using PYTHIA8~\cite{Sjostrand:2007gs},
and the fast detector simulation was carried out with
DELPHES~\cite{deFavereau:2013fsa}. Jet analyses were performed using
FASTJET~\cite{Cacciari:2011ma}. Exclusively for the presently
available ATLAS NC data~\cite{ATLAS:2019erb}, the detector response
was simulated using Rivet~\cite{Bierlich:2019rhm, Buckley:2019stt},
with the analysis code provided by the experimental
collaboration. \smallskip

In the analysis of NC (CC) DY process we employed the invariant
(transverse) mass distribution to perform the USMEFT fit to the
corresponding pseudo-data. The invariant (transverse) mass binning
used is from 0.5 (1.0) TeV to 3.5 TeV and an overflow bin for all
events with invariant (transverse) masses above 3.5 TeV.  Defining as
$N^\text{P-D,NC(CC)}_i$ the pseudo-data in bin $i$ for the NC (CC)
process for a given BSM model and parameters, and
$N_i^\text{EFT,NC(CC)}$ the expected number of events in that same bin
for a given value of USMEFT WC, the chi-squared functions for the DY
distributions are constructed as
\begin{equation}
  \chi^2_{\rm DY,NC(CC)} (M,\beta|\vec c_{\rm NC(CC)}) = \sum_i \dfrac{\left[N^\text{P-D,NC(CC)}_i (M,\beta)
      -N_i^\text{EFT,NC(CC)} (\vec c_{\rm NC(CC)}) \right]^2}
  {N^\text{P-D,NC(CC)}_i (M,\beta)+ \delta_{\text{syst},\, i}^2}\,,
\end{equation}
where
\begin{align}
    N^\text{P-D,NC(CC)}_i(M,\beta) &= {\cal L} \,\sigma^\text{P-D,NC(CC)}_i(M,\beta)\,,\\
    N^\text{EFT,NC(CC)}_i (\vec c_{\rm NC(CC)}) &= {\cal L} \,\sigma^\text{EFT,NC(CC)}_i (\vec c_{\rm NC(CC)})\,,\\
    \delta_{\text{syst},\, i} &= 0.015 \times {m_i}\, [\text{TeV}] \times N^\text{PD,NC(CC)}_i\, .
\label{eq:dyunc}
\end{align}
${\cal L}$ denotes the luminosity, which we take to be
$3000~\text{fb}^{-1}$,  while
  $\sigma^\text{P-D,NC(CC)}_i (M,\beta)$ stands for the BSM cross
  section for model parameters $(M,\beta)$ in that bin and
$\sigma^\text{EFT,NC(CC)}_i(\vec c_{\rm NC(CC)})$ is the corresponding
predicted cross section within USMEFT for WC
$$\vec c_{\rm NC}\equiv\left(\overline{c}_{BW},
  \overline{c}_{\Phi,1},\overline{\Delta}_{4F},
  \overline{c}_{2JB}, c^{(2)}_{\psi^4 D^2},
  c^{(3)}_{\psi^4 D^2}, c^{(7)}_{\psi^4 H^2}\right)\, , $$
and   
$$\vec c_{\rm CC}\equiv\left({ \overline{c}_{BW}}, \overline{c}_{\Phi,1},\overline{\Delta}_{4F},
\overline{c}^{(3)}_{W^2H^4}, c^{(3)}_{\psi^4 D^2} \right)\;. $$
Both cross sections include the detector effects and event selection
cuts. Also following Ref.~\cite{ATLAS:2018tvr}, we have included a
systematic uncertainty of 1.5\% $m_i$ (TeV) to simulate energy
resolution~\cite{ATLAS:2018qvs}. $m_i$ is the value of the
corresponding invariant (transverse) mass in the center of the bin $i$
for all bins but for the overflow bin for which we take as $m_i$ the
value of the mass in the lower edge of the bin.  \smallskip

In the EWPO analyses, we include 12 $Z$-pole
observables~\cite{ALEPH:2005ab}: $\Gamma_Z$, $\sigma_{h}^{0}$,
${\cal A}_{\ell}(\tau^{\rm pol})$, $R^0_\ell $,
${\cal A}_{\ell}({\rm SLD})$, $A_{\rm FB}^{0,l}$ $R^0_c$, $R^0_b$,
${\cal A}_{c}$,${\cal A}_{b}$, $A_{\rm FB}^{0,c}$, and
$A_{\rm FB}^{0,b}$ and two $W$ pole observables $M_W$ and $\Gamma_W$
taken from~\cite{ParticleDataGroup:2024cfk, ALEPH:2005ab}. The SM
predictions and their uncertainties due to variations of the SM
parameters were extracted from~\cite{deBlas:2022hdk}. In order to
eliminate the fluctuations associated with the central values of the
real EWPO results, in our forecasts, we simulate the central values of
EWPOs following the predictions of the BSM model for the given set of
parameters ($M,\beta$).  With those we define a function
$\chi^2_\text{EWPO} (M,\beta|\vec c_{\rm EWPO})$ which depends on the
WC:
$$\vec c_{\rm EWPO}\equiv \left(\overline{c}_{BW} , \overline{c}_{\Phi,1},\overline{\Delta}_{4F},
\overline{c}^{(3)}_{W^2H^4},c^{(2)}_{\psi^4 D^2}, c^{(3)}_{\psi^4 D^2}\right) \; .$$
In fact, EWPO data is capable of constraining only 4 combinations of
these parameters at order $1/\Lambda^4$~\cite{Corbett:2025oqk},
however the combination of the EWPO and DY data resolves the blind
directions of the EWPO. \smallskip

As mentioned in the introduction, we perform our USMEFT fits in four
different variants of the analysis:
\begin{itemize}
\item In the first case, we truncate the expansion at
  ${\cal O}(\Lambda^{-2})$ by considering only the first two terms in
  Eq.~\eqref{eq:ampli}. In this scenario referred as $(d=6)$ analysis,
  we include only the four coefficients of the dimension-six operators
  and keep their effects in the observables at linear order in the
  coefficients.
\item In the second analysis, we include all contributions in
  Eq.~\eqref{eq:ampli} but the last term. This accounts to consider
  only the four coefficients of the dimension-six operators and to
  keep their effects in the observables up to quadratic order in their
  WC.  We refer to this case as $(d=6)^2$ analysis.
\item In the third case, we perform a full
  ${\cal O}(\Lambda^{-4})$ analysis by considering all terms in
  Eq.~\eqref{eq:ampli}. This case, dubbed $(d=8)$ analysis, contains
  all the contributions in the previous scenario as well as linear one
  coming from dimension-eight operators.  
\item The fourth analysis is the same as the third  but employing the
  two coefficient combinations in Eq.~\eqref{eq:comb}. We refer to
  this case as $(d=8)'$ analysis.
\end{itemize}  

The last ingredient in our studies, which serves the purpose of
answering our first question on the HL-LHC potential to uncover NP
signals using the EFT framework, we defined for each WC, $c_j$, the
quantity
\begin{equation}
  f_{\rm NP} (c_j) \equiv \frac{ | c^{\rm best\,fit}_j|}{\Delta c_j} 
\label{eq:fnp}
\end{equation}
where $\Delta c_j$ is half of the difference between the upper and
lower 68\% CL limits on $c_j$. This quantity measures the deviation
from the SM prediction ($c_j=0$) in number of $1\sigma$
  deviations.  \smallskip

\section{Results}
\label{sec:results}

\begin{figure}[!t]
 \includegraphics[width=0.8\textwidth]{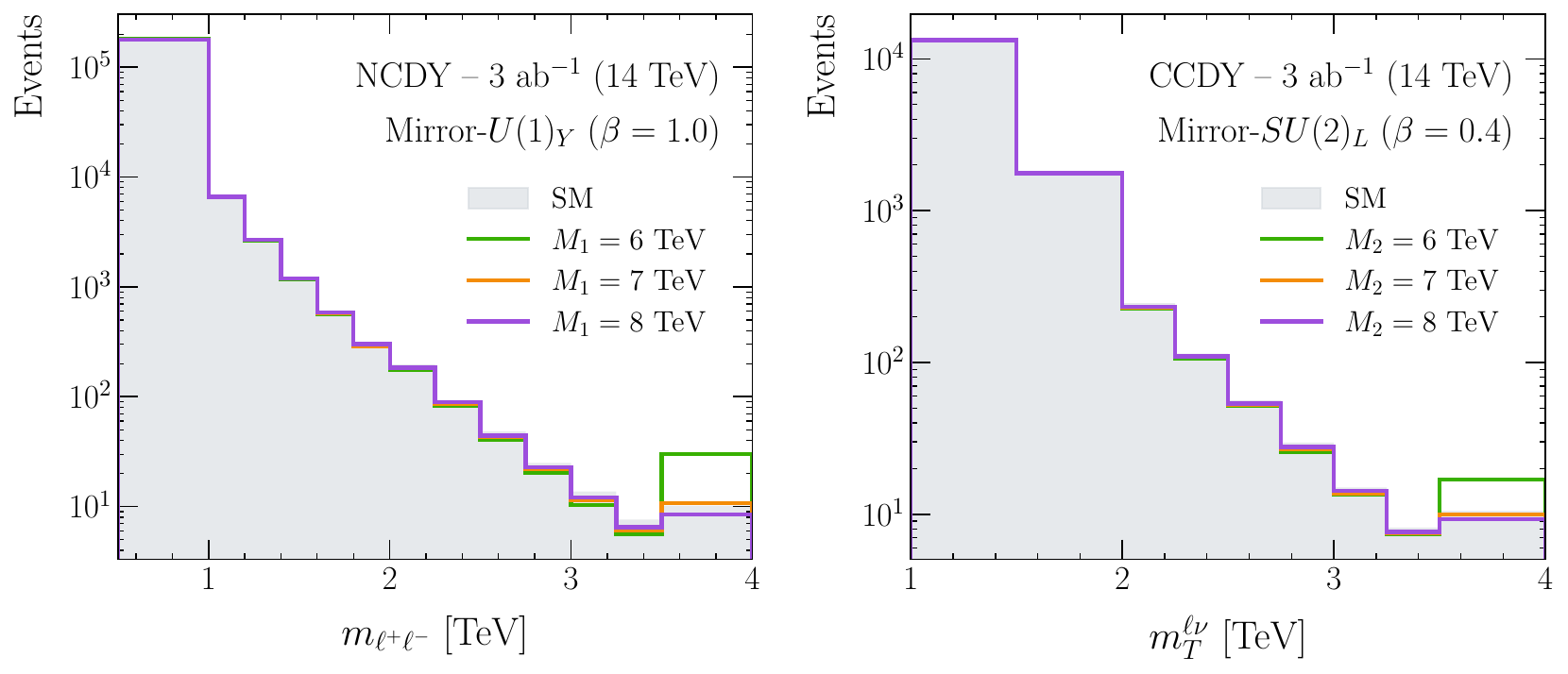}\hspace*{-1cm}
 \caption {The  left (right) panel displays the expect number
   of events distribution as a function of the invariant (transverse)
   mass for the mirror-$U(1)_Y$ (mirror-$SU(2)_L$) extension of the
   SM.  We assumed an integrated luminosity of 3 ab$^{-1}$.}
  \label{fig:dist}
\end{figure}

Our starting point is the generation of the pseudo-data for the
invariant (transverse) mass distribution for DY NC (CC).  We generated
DY pseudo-data for gauge boson masses between 4 and 10 TeV and several
values of the coupling scaling $\beta$.  We show in
Fig.~\ref{fig:dist} examples of these distributions for NC pseudo-data
for the mirror-$U(1)_Y$ (mirror-$SU(2)_L$) model in the left
  (right) panel for several values of the model parameters.  As a
starting step, taking into account that the EFT contribution grows at
high invariant (transverse) masses, as illustrated in the figure, we
obtained the range of model parameters where a simple counting
experiment using the events in the overflow bin is enough to guarantee
a $5\sigma$ discovery of NP. We do so because, certainly, in this
parameter region a new rebinning or a resonance search would be more
efficient than an EFT analysis to confirm and identify the NP
states. \smallskip

\begin{figure}[!t]\centering
  \includegraphics[width=0.9\textwidth,height=0.75\textheight]{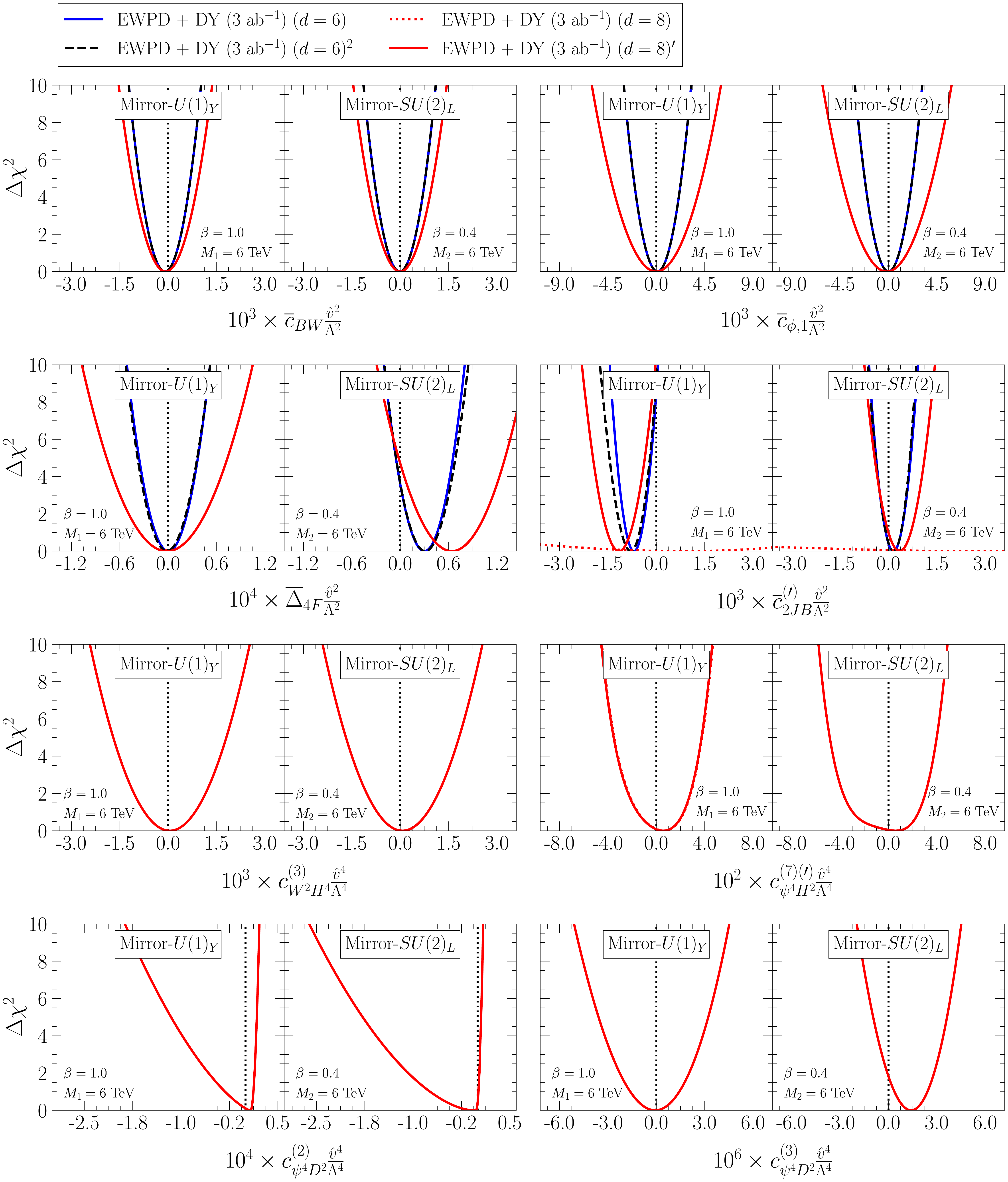}\hspace*{-1cm}
 \caption {One-dimensional projections of $\Delta \chi^2$ after
   marginalization over all other Wilson coefficients for the different
   analyses performed as labeled in the figure. Notice that for all coefficients but
   $\overline{c}_{2JB}$ and $c^{(7)}_{\psi^4H^2}$ the curves for the analysis $(d=8)$
  and $(d=8)'$ fully overlap so only the full red lines are seen (see text for details).}
  \label{fig:1d}
\end{figure}

Next, for each value of the model parameters for which we have
generated the pseudo-data we performed an EFT fit of the DY
pseudo-data and EWPO without imposing any assumption on the eight WCs
involved. From each of the analyses we find the best fit value and
uncertainties of the eight WCs.  An example is shown in
Fig.~\ref{fig:1d}, where we plot the one-dimension projections of
$\Delta \chi^2$ for different orders considered in the analysis for one
example of pseudo-data generated for each of the models.  To guide the
eye, we plot as a vertical dotted line the SM value ``0'' of the
coefficients.\smallskip

\begin{figure}[!t]
 \includegraphics[width=0.8\textwidth]{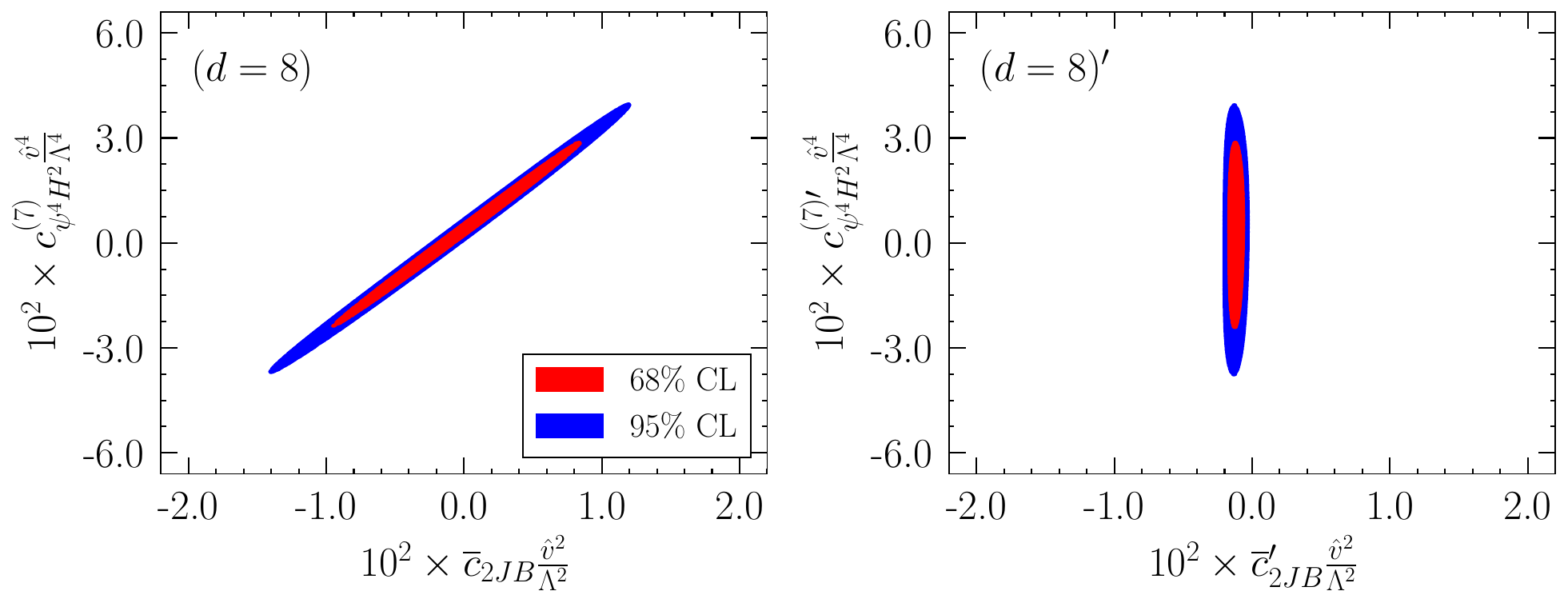}\hspace*{-1cm}
 \caption {{\bf Left}:Allowed regions for the parameters
   $\overline{c}_{2JB}$ and $c^{(7)}_{\psi^4H^2}$ for the fit to the
   pseudo-data for mirror-$U(1)_Y$ model with $M_1=1$ TeV and
   $\beta=1$ at order $(d=8)$ after marginalization over all other
   parameters.  {\bf Right}: Same region for the combinations
   $\overline{c}_{2JB}^\prime$ and $c^{(7)\prime}_{\psi^4H^2}$ in
   Eq.~\eqref{eq:comb}.}
 \label{fig:corr}
\end{figure}

Focusing on the result obtained at $(d=6)$ (blue lines) and $(d=6)^2$
(black dashed lines), we see that, both analyses lead to similar
results indicating that the bounds are dominated by the dimension-six
cross section. In particular we see that the coefficients for which
the best fit value is most prominently away from zero at this order are
$\overline{c}_{2JB}$ for the fit with the mirror-$U(1)_Y$ pseudo-data
and $\overline{\Delta}_{4F}$ ({\em i.e.}  $\overline{c}_{2JW}$) for
the mirror-$SU(2)_L$ pseudo-data.  This is expected because these are
the coefficients generated by direct matching at tree level (see Table
~\ref{tab:WC}).  Furthermore the four-vertex contact interactions
associated with $\bar{c}^{(\prime)}_{2JB}$ and
$\bar{c}_{2JW}$ presents the fastest growth with subprocess
center-of-mass energy.  The other minor deviations arise from the
smaller contributions to indirect finite-renormalization effects, or
by correlations between the determined coefficients in the fit.
Precisely on that respect, one can see that when including also the
dimension-eight operators the sensitivity to the characteristic
non-zero value of $\overline{c}_{2JB}$ for the fit with the
mirror-$U(1)_Y$ pseudo-data is lost: the corresponding $\Delta\chi^2$
is the red dotted line at the bottom of those panels which is almost
flat zero within the plotted range.  The reason for this loss of
sensitivity is that, as shown in Ref.~\cite{Corbett:2025oqk}, the
analysis at $1/\Lambda^4$ order presents a strong correlation between
$\bar{c}_{2JB}$ and $c^{(7)}_{\psi^4H^2}$ limiting our ability to
precisely determine both of the WCs simultaneously. We see this in the
left panel of Fig.~\ref{fig:corr} where we plot their allowed regions
which, as seen on the figure, display a very strong
correlation.\smallskip

These two operators enter DY in the four-fermion NC contact amplitudes
dominantly in the linear combination
$\overline{c}_{2JB}\frac{\vh^2}{\Lambda^2}-\frac{\sh}{2\ch}
c^{(7)}_{\psi^4 H^2} \frac{\vh^4}{\Lambda^4}$
leading to the very
strong correlation observed and the substantial weakening of the
bounds on $c_{2JB}$ when compared to the analysis performed at $(d=6)$
or $(d=6)^2$. However, it is important to notice that this correlation
appears in the USMEFT prediction irrespective of the pseudo-data
fitted.  This allows to perform our analysis {\sl blindly}, this is,
deciding a priori to use a different basis for those two coefficients
in which one of the new coefficients is chosen to be the combination
that dominantly contributes to DY and the other the orthogonal one.
To this end we define the ``prime'' combinations
\begin{eqnarray}
  \bar{c}^\prime_{2JB}
  ~\frac{\hat{v}^2}{\Lambda^2} &=&
     \bar{c}_{2JB}~\frac{\hat{v}^2}{\Lambda^2}
        + 0.33 ~ c^{(7)}_{\psi^4H^2}~\frac{\hat{v}^4}{\Lambda^4} \;,
\nonumber
  \\
&& \label{eq:comb}
  \\
  c^{(7)\, \prime}_{\psi^4H^2} ~\frac{\hat{v}^4}{\Lambda^4}&=& c^{(7)}_{\psi^4H^2}~\frac{\hat{v}^4}{\Lambda^4}  - 0.33~\bar{c}_{2JB} ~\frac{\hat{v}^2}{\Lambda^2}\;.
\nonumber
\end{eqnarray}
For all other operator coefficients the prime and unprimed coefficients are the same.\smallskip

As seen in the right panel in Fig.~\ref{fig:corr} in these variables
the correlation disappears and consequently the coefficient
$\bar{c}^\prime_{2JB}$ can be more precisely determined as can be seen
also in the one-dimensional projection for this variable for the
$(d=8)'$ case in Fig.~\ref{fig:1d} (full red lines).  With all this we
find that including the dimension-8 operators, still allows for the
identification of the operator coefficients with dominant NP effects,
$\bar{c}^\prime_{2JB}$ for mirror-$U(1)_Y$ pseudo-data fits, and
$\bar{c}_{2JW}$ for mirror-$SU(2)_L$ pseudo-data fits while the
precision in the determination of these coefficients is only mildly
affected despite the larger parameters space. \smallskip

\begin{figure}[!t]
  \includegraphics[width=0.7\textwidth]{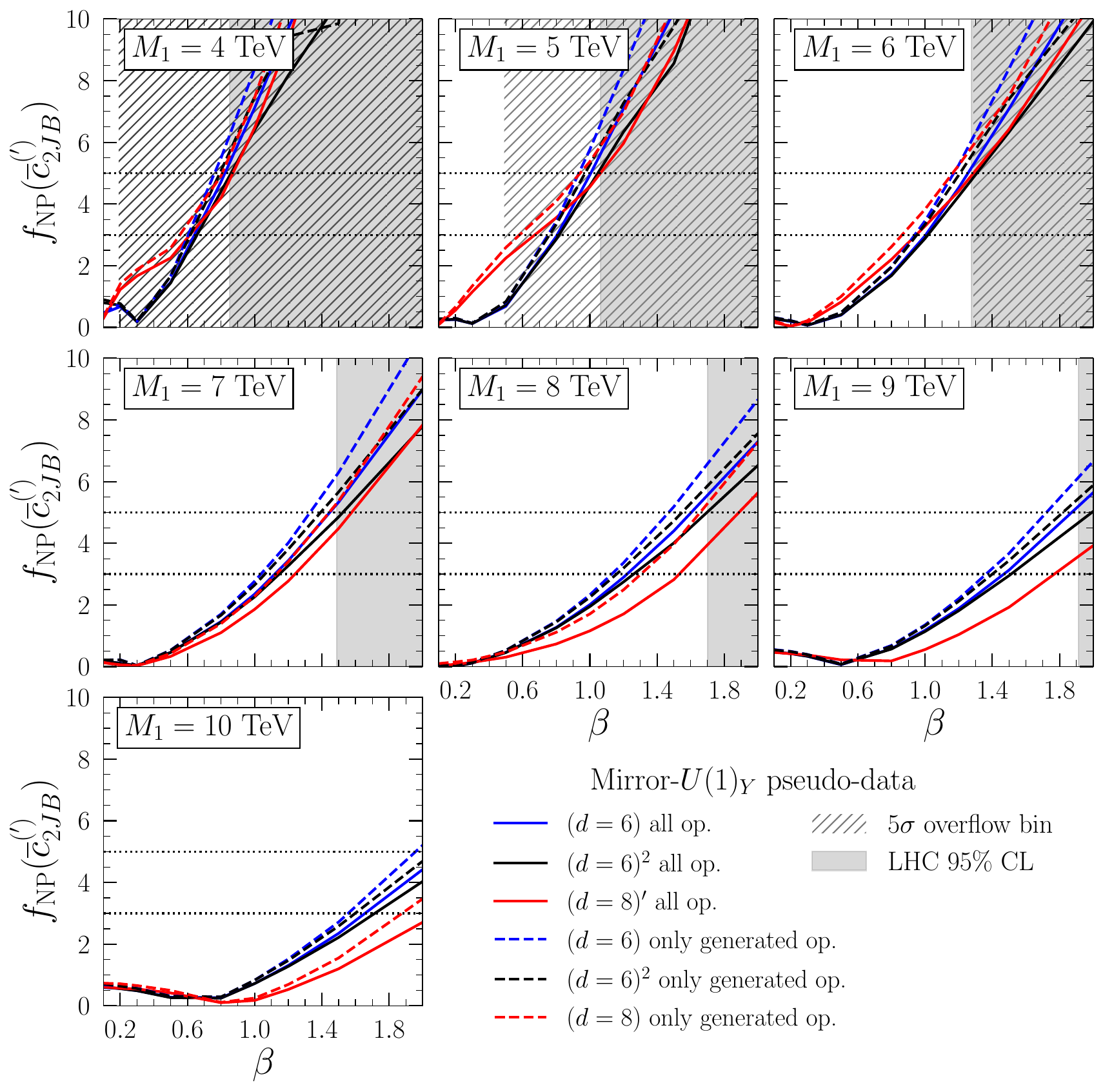}
  \vglue -0.3cm
  \caption {$f_{\rm NP}$ as a function of $\beta$ for
    $c^{(\prime)}_{2JB}$ obtained with  fit to pseudo-data generated with the
    mirror-$U(1)_Y$ model for the masses as labeled in the figure.
    The full lines correspond to the analysis performed
    with all the eight relevant operator coefficients, while the dashed lines
    correspond to the analysis performed including only coefficients
    of the operators generated in the model
    (three at $(d=6)^{(2)}$
    and five at $(d=8)'$).
    The shaded area is excluded at 95\% CL by
    the present available data while the hashed one is the one where
    the overflow alone leads to a 5$\sigma$ discovery. We present the
    results for different $1/\Lambda$ orders as indicated.
    We assumed
    an integrated luminosity of 3 ab$^{-1}$.}
      \label{fig:u1xfnp}
\end{figure}
\begin{figure}[!t]
  \includegraphics[width=0.7\textwidth]{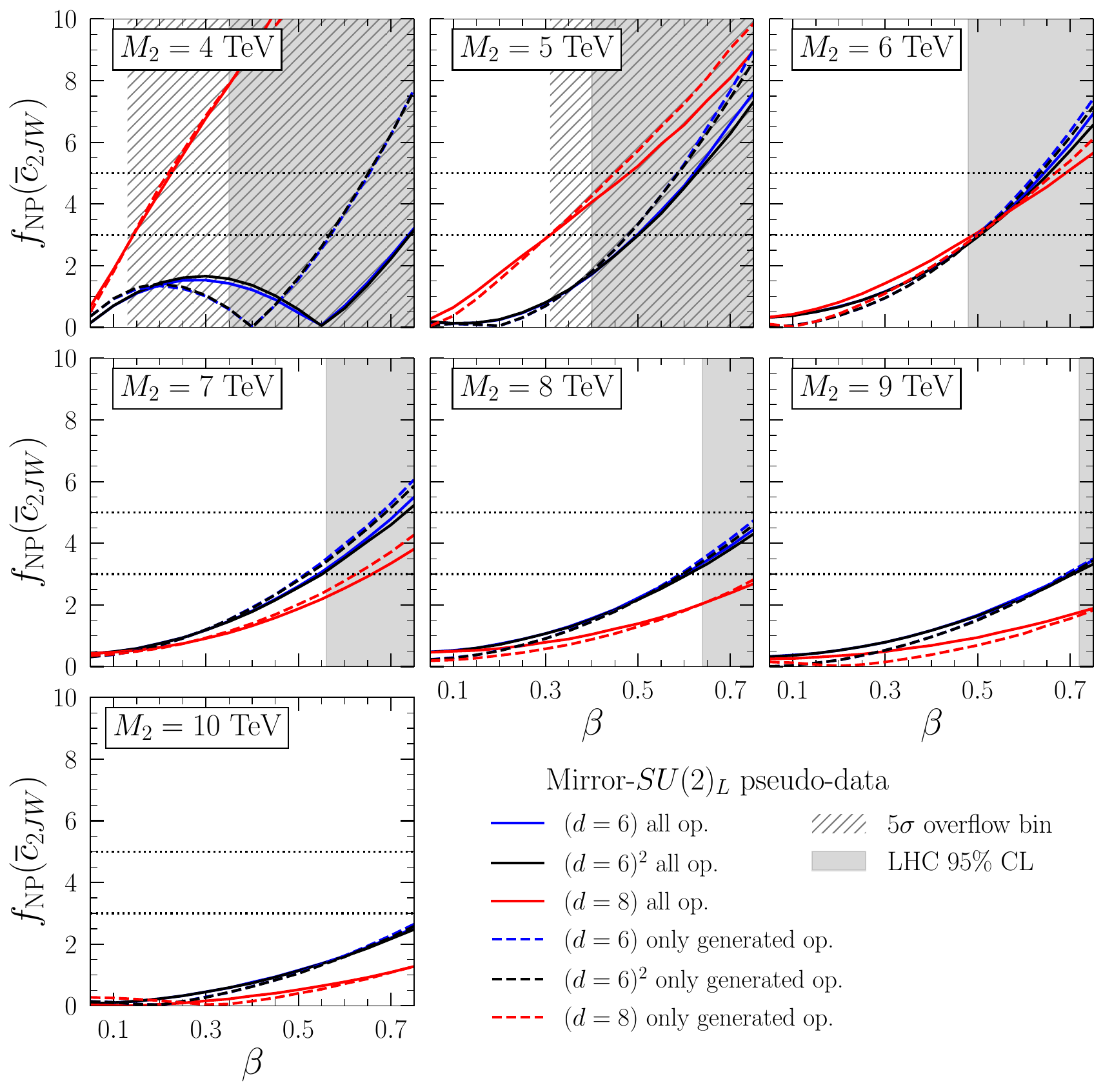}
  \vglue -0.3cm
  \caption {$f_{\rm NP}(\bar{c}_{2JW})$ as a function of $\beta$
    obtained with fit to pseudo-data generated with the
    mirror-$SU(2)_L$ model for the masses as labeled in the figure.
    Conventions are the same as in Fig.~\ref{fig:u1xfnp}. In this case
    the dashed lines correspond to fits including the coefficients of
    the operators which would be generated in the mirror-$SU(2)_L$
    model, (two at $(d=6)^{(2)}$ and three at $(d=8)'$).}
  \label{fig:su2xfnp}
\end{figure}

From Fig.~\ref{fig:1d}, we also anticipate that, for any point in the 
model parameter space, those two couplings are the most sensitive to the 
NP measure $f_{\rm NP}$. As mentioned above, this is expected since the 
four-point contact interactions associated with
$\bar{c}^{(\prime)}_{2JB}$ and $\bar{c}_{2JW}$
exhibit the fastest growth with the subprocess center-of-mass energy. 
In principle, as shown in Table~\ref{tab:WC},
$c^{(2)}_{\psi^4D^2}$ and $c^{(3)}_{\psi^4D^2}$ are 
also generated through tree-level matching of direct effects in
the mirror-$U(1)_Y$ and mirror-$SU(2)_L$ 
models at dimension-eight. Despite inducing energy-growing effects,
they are poorly constrained because of the stronger
mass suppression.
\smallskip

Figures~\ref{fig:u1xfnp} and ~\ref{fig:su2xfnp} depict, respectively,
the dependence of $f_{\rm NP}(\bar{c}^{(\prime)}_{2JB})$ and
$f_{\rm NP}(\bar{c}_{2JW})$ on $\beta$ for different values of $M_1$
and $M_2$. First, as expected, the simple analysis of the overflow bin
(hashed area) excludes a large fraction of the $\beta$ range for the
smaller mass values  (4 and 5 TeV) allowing only small values of
$\beta$.  For the mirror-$U(1)_Y$ case for these lighter masses, the
EFT analyses do not lead to any evidence of NP due to the small value
of $\beta$, while for the mirror-$SU(2)_L$ case the sensitivity is a
bit higher but still marginal.  In addition, as we can see, that for
the mirror-$SU(2)_L$ the presently available DY data exclude
a larger fraction of the parameter space, since this model contributes
to both NC and CC DY processes, especially for masses $M\lesssim 5$
TeV.  \smallskip

The different curves for $f_{\rm NP}$  are obtained with  the analyses
performed at different orders in the $1/\Lambda$
expansion. Furthermore, to test the robustness of the conclusions we
have performed two variants of the analysis. One in which all the
operator coefficients of a given order are included (solid lines), and
one in which we artificially fix to zero the coefficients which we
know that are not generated by the model (dashed lines).  In this
case, we only fit three dimension-six coefficients --
$\overline{c}_{BW}$,$\overline{c}_{\Phi,1}$,
$\overline{c}_{2JB}^{(')}$ -- and two additional dimension-eight
coefficients --$c^{(2)}_{\psi^4 D^2}$ and
$\overline{c}^{(3)}_{W^2H^4}$ -- in the fits to the mirror-$U(1)_Y$
pseudo-data shown in Fig.~\ref{fig:u1xfnp}, and two dimension-six
coefficients -- $\overline{c}_{BW}$, and $\overline{c}_{2JW}$ -- and
one additional dimension-eight coefficient --$c^{(3)}_{\psi^4 D^2}$--
in the fits to mirror-$SU(2)_L$ pseudo-data shown in
Fig.~\ref{fig:su2xfnp}. \smallskip

Comparing the results obtained at the different orders in the operator
expansion, we see in Fig.~\ref{fig:u1xfnp} that for the fits to
mirror-$U(1)_Y$ pseudo-data the $d=6$ and $(d=6)^2$ results are rather
similar and the $(d=8)'$ analyses reduces $f_{\rm NP}$ by about
10--20\% for higher $M_1$ masses. Altogether, we find that an EFT
$5\sigma$ signal ($3\sigma$ evidence) can be observed for $M_1$
between 6 and 8 (9) TeV and $\beta$ is in the range 1.3 to 1.9 (1 to
1.8). \smallskip

Similarly, in the mirror-$SU(2)_L$ scenario, we find that the inclusion of
the full dimension-eight contributions reduces the reach for NP due to the
slight increase of the 68\% allowed range of $\bar{c}_{2JW}$.  In this
case we find that $5\sigma$ discovery using the EFT framework is
possible for $M_2$ ranging from 5 to 10 TeV and $\beta$ between 0.5
and 1.5 when we consider the $1/\Lambda^4$ contributions. A
$3\sigma$ evidence of NP is possible for these masses  for
$\beta$ ranging from 0.3 to 1.2. \smallskip

Finally, comparing the solid and dashed lines in both figures we find
that model parameters for which the EFT analysis is consistent, the
sensitivity to NP is not improved by artificially selecting only the operators 
generated by the model. 
\smallskip 

\begin{figure}[!p]
  \includegraphics[width=0.7\textwidth]{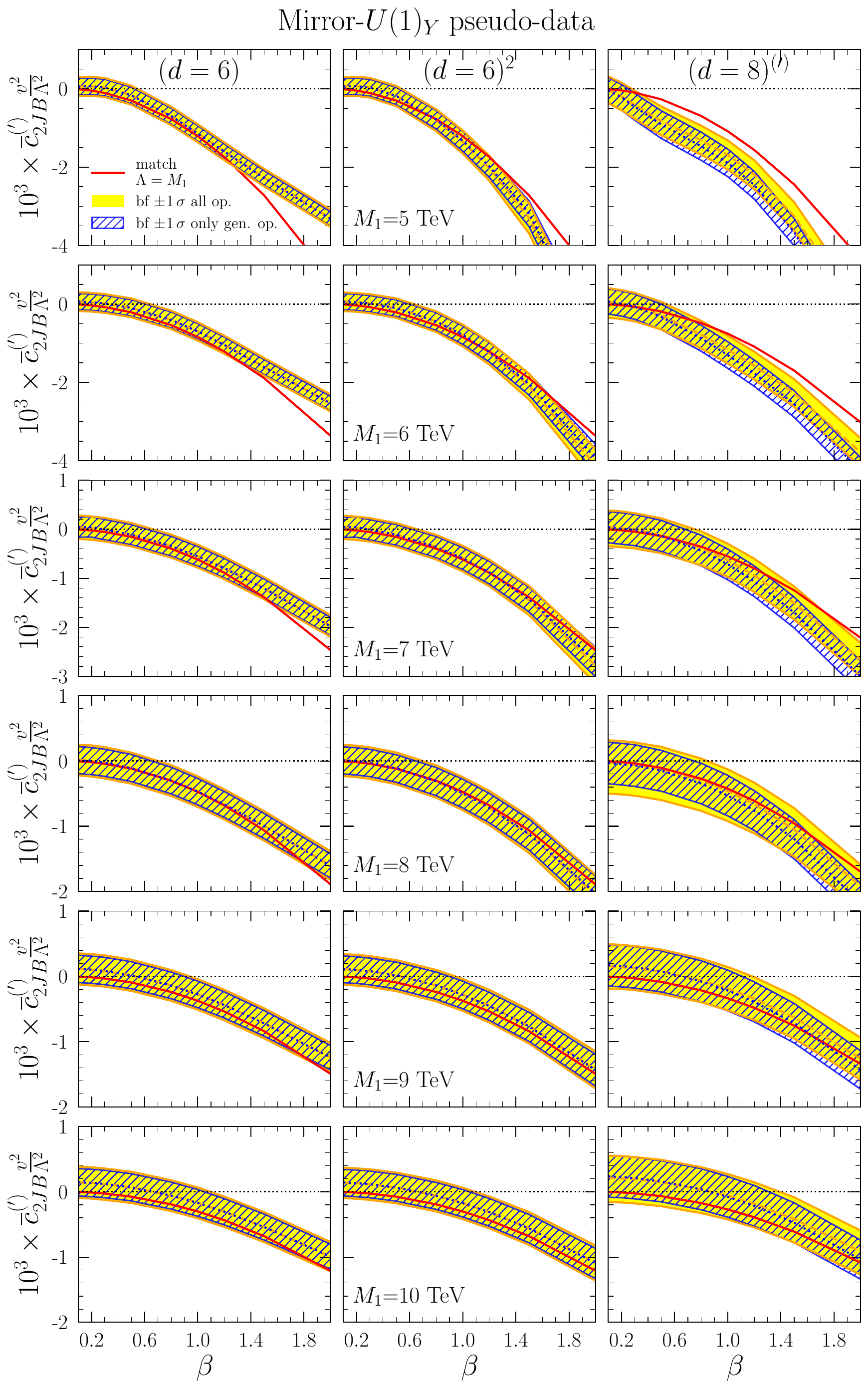}
  \caption{Allowed one sigma regions for $c^{(\prime)}_{2JB}$ as a
    function of $\beta$ obtained with  fit to pseudo-data generated with
    mirror-$U(1)_Y$ model for the masses as labeled in the figure.
    The three columns correspond to the analysis performed at
    different orders in the operator expansion. In each panel we show
    as yellow band the 1$\sigma$ allowed region around the best fit
    (orange dashed line)
    obtained with the analysis including all operators in that order of
    the expansion. The blue hatched band  correspond to the 1$\sigma$ allowed
    region around the best fit (blue dotted line)
    obtained with the analysis including only the operators which are generated
    in the model at that order. The red line is the predicted value
    by the matching (see Tab.~\ref{tab:WC}).
    We assumed
    an integrated luminosity of 3 ab$^{-1}$. }
  \label{fig:u1xmat}
\end{figure}
\newpage
\begin{figure}[h!]
  \includegraphics[width=0.7\textwidth]{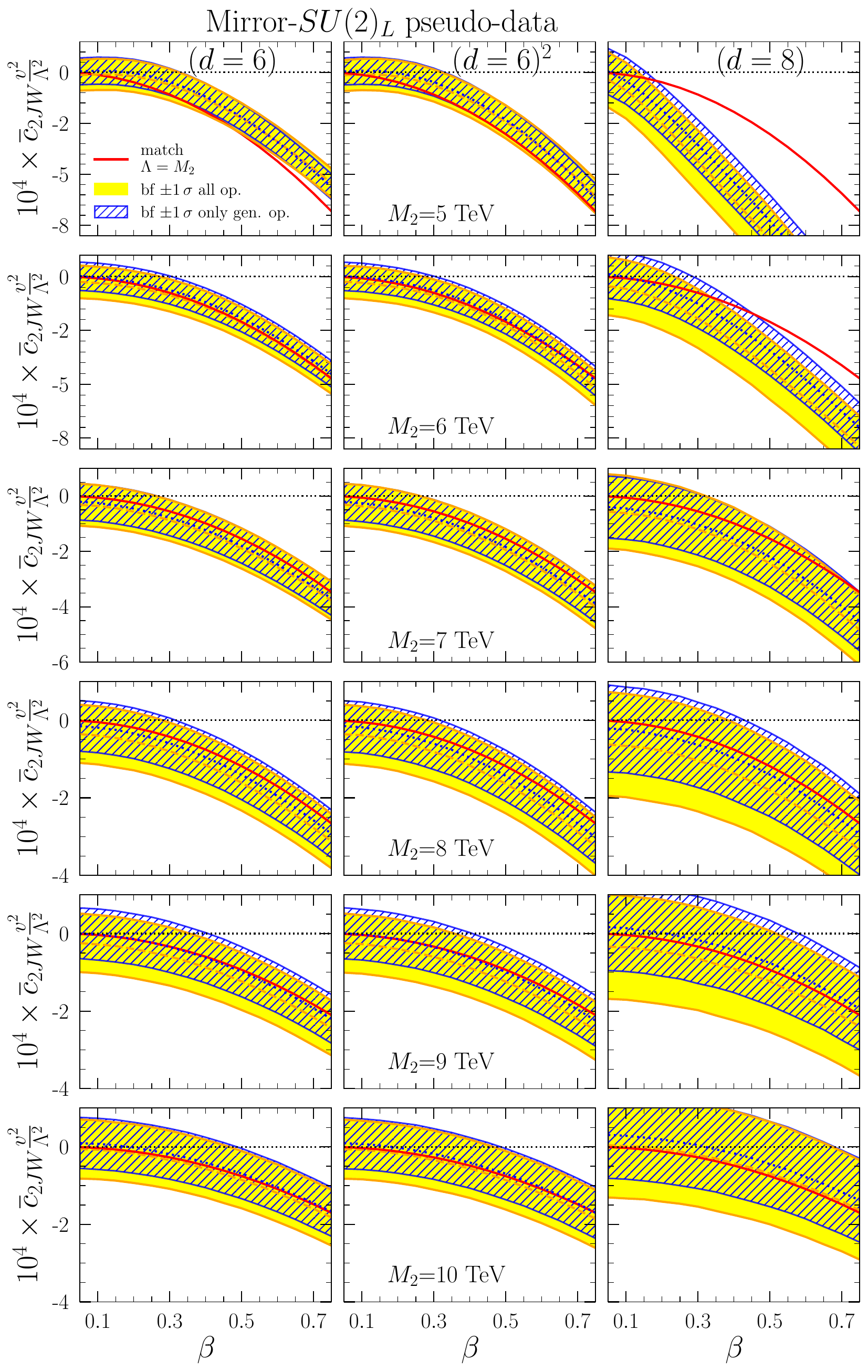}
  \caption{Allowed one sigma regions for $c_{2JW}$ as a
    function of $\beta$ obtained with  fit to pseudo-data generated with the
    mirror-$SU(2)_L$ model for the masses as labeled in the figure.
    Conventions are the same as in Fig.~\ref{fig:u1xmat}.}
    \label{fig:su2xmat}
\end{figure}

We now turn to the question on whether the extracted values of the
WC's capture the properties of the BSM in each case. To address this
issue, we present in Figs.~\ref{fig:u1xmat} and
Figs.~\ref{fig:su2xmat} the $1\sigma$ allowed band of
$\overline{c}^{(\prime)}_{2JB}$ extracted from the fits to
mirror-$U(1)$ pseudo-data and $\overline{c}_{2JW}$ from the
fits to mirror-$SU(2)$ pseudo-data respectively. These results are
shown as function of the coupling $\beta$ with the rows corresponding
to a given mass and the columns to the different orders considered in
the fit.  In each panel the red solid line represents the predicted
value of the WC from the matching in the model at that order.
Furthermore, as for the $f_{\rm NP}$ figures, to test the robustness
of the conclusions, we show the results for the complete fits to each
order, as well as the result of the ``biased'' fit in which only the
subset of known-to-be-generated operator coefficients are
included.\smallskip

These result for the fits to both models' pseudo-data present similar
behaviors. First of all, for heavier $M_{1,2}$ masses ($\gtrsim 7$ TeV) the
allowed $1\sigma$ regions of $\bar{c}^{(\prime)}_{2JB}$
and $\bar{c}_{2JW}$ contain the
{\sl true} value of this WC (i.e. that predicted by the matching).
Moreover, the results are stable with respect
to the $1/\Lambda$ order indicating that the EFT is describing well
the low-energy data. As we consider smaller masses the global
fit $1\sigma$ bands do not contain the true  values for large $\beta$,
suggesting that the EFT is being employed beyond its
applicability range due to the proximity and size of the resonance.
The inclusion of higher orders does not mitigate this departure,
possibly indicating that the effective extension of the window of applicability 
of the $1/\Lambda^4$ and higher order contributions is narrow in comparison
with that of the $1/\Lambda^2$ terms.  \smallskip

Finally, comparing the blue hatched and solid yellow bands we see that
the behavior does not change significantly in the restricted fits.
Thus altogether, we find that our results are robust even when
blindly fitting the data with minimal theoretical bias. \smallskip

\section{Summary}
\label{sec:summary}

The expected accumulated luminosity at the HL-LHC will allow detailed
analyses of the high energy tail of kinematic distributions. In the
absence of new resonances, the EFT framework for the analysis of those
distributions is a powerful tool to search for signs of NP as well as
to constrain possible extensions of the SM . In this work, we study
the NC- and CC-DY processes to assess the HL-LHC potential to search
for NP using USMEFT as the analysis tool. To this end, we simulated
the expected signal of two universal SM extensions: the
mirror-$U(1)_Y$ and mirror-$SU(2)_L$ and performed a fit of that
pseudo-data for the NC and CC DY distributions together with the EWPO
with the minimal theoretical priors at different orders in the
operator truncation. \smallskip

We found that the analyses performed at orders dimension-six and
dimension-six-squared lead to similar results due to the tight limits
on the dimension-six Wilson coefficients. The inclusion of
dimension-eight operators in the analyses lead to slightly less
precise results due to the increased number of four-fermion operator
coefficients. \smallskip

More qualitatively, our unbiased fits of the pseudo-data for both
models show that the HL-LHC will be able to lead to a $5\sigma$
discovery ($3\sigma$ evidence) for masses up to $\simeq 7$ (9) TeV
provided the coupling is of similar strength to the corresponding the
SM $U(1)_Y$ and $SU(2)_L$ couplings.  We also analyzed how well a
global fit with arbitrary WCs can capture the true value of the BSM
theory. Except for the parameter regions where the new resonance could
be directly observed, the dominant WC ($\bar{c}^{(\prime)}_{2JB}$ and
$\bar{c}_{2JW}$) true model value is inside the $1\sigma$ allowed
region for these WCs for a large range of masses and
couplings.\smallskip

We also find that that both the NP reach as well as the identification
of the BSM model obtained in the most unbiased fit including all the WCs
at a given order of the truncation are very similar to those obtained
imposing the relations between WCs generated in each of the models.
This further strengthens the robustness of our conclusion: fits with
the minimal theoretical bias can be used not only to unveil the
existence of new physics but also to accurately extract its
properties. \smallskip

\acknowledgments

OJPE is partially supported by CNPq grant number 302120/2025-4.
M.M. is supported by FAPESP grant 2022/11293-8.  This project
is funded by USA-NSF grant PHY-25104424.  It has also received support
from the European Union's Horizon Europe research and innovation
programme under the Marie Sk\l odowska-Curie Staff Exchange grant
agreement No 101086085 -- ASYMMETRY''.  It also receives support from
grants PID2019-105614GB-C21, and ``Unit of Excellence Maria de
Maeztu'' award to the ICC-UB CEX2024-001451-M funded by
MICIU/AEI/10.13039/501100011033.

\bibliography{references}

\newpage
\appendix

\section{Effective interactions entering our analyses}
\label{app:couplings}

There are six dimension-six and eleven dimension-eight operators
entering the Drell-Yan production in USMEFT.  In order to perform the
Monte Carlo event simulations it is more convenient to work with the
operators in the rotated basis including in addition the indirect
effects induced by the finite renormalization of the SM
parameters~\cite{Corbett:2023qtg}.  \smallskip

Here, we chose the input parameters to be
$\{ \widehat{\alpha}_{\rm em} \;,\; \widehat{G}_F \;,\;
\widehat{M}_Z\}$ and take into account the following three relations
to define the renormalized parameters
\begin{equation}
  \hat{e} = \sqrt{ 4 \pi \hat{\alpha}_{\rm em}} \;\;\;,\;\;\;
  \hat{v}^2 = \frac{1}{\sqrt{2} \, \hat{G}_F} 
     \label{eq:vhat}
     \;\;\;\hbox{and}\;\;\;
     \hat{c}^2 \hat{s}^2 = \frac{\pi \widehat{\alpha}_{\rm em}}{\sqrt{2}\,
     \widehat{G}_F \widehat{M}^2_Z} \;,
     \nonumber
\end{equation}
where $\sh$ ($\ch$) is the sine (cosine) of the weak mixing angle
$\hat{\theta}$. The contributions
of fermionic operators to the muon decay width are written as
\begin{equation}
  \left[2\langle H^\dagger H \rangle - \frac{1}{\sqrt{2} \hat{G}_F}
  \right]_{\rm fermionic}\equiv  \frac{\vh^4}{\Lambda^2} \Delta_{4F}
  + \frac{\hat{v}^6}{\Lambda^4} \Delta^{(8)}_{4F}  \;,
\label{eq:vt}
\end{equation}
where $\Delta_{4F}$ ($\Delta_{4F}^{(8)}$) contains the dimension-six
(-eight) contributions.
%
After finite renormalization of the SM inputs and accounting for the
direct contribution from the fermionic dimension-eight operators
containing two fermion fields we can parametrize the $Z$ coupling to
fermion pairs $\bar{f}f$ as
\begin{equation}
  \frac{\eh}{\sh\ch} ~\left[\hat{g}_{L,R}^f\, \left(1+\Delta \overline{g}^\prime
    +\Delta g^{\prime\,\square}+\frac{p^2}{\widehat{M}_Z^2}\Delta g^\prime\right)
    +Q^f\, \left(\Delta \overline{g}
    +\Delta g^\square+\frac{p^2}{\widehat{M}_Z^2}\Delta g'\right)\right]\;,
\label{eq:dgZ}
\end{equation}
and the $W$ coupling to left-handed fermions as
\begin{equation}  
 \frac{1}{\sqrt{2}} \frac{\eh}{\sh} ~
\left(1+\Delta \overline{g}_W
    +\Delta g_W^\square+\frac{p^2}{\widehat{M}_W^2}\Delta g_W'\right)\;,
\label{eq:dgW}  
\end{equation}
where $\hat g^f_L=T_3^f-\sh^2 Q^f$, $\hat g^f_R=-\sh^2 Q^f$, $T_3^f$
is the fermion's third component of isospin, and $Q^f$ is its
charge. In the expressions above $p^2$ is the square of the
four-momentum of the corresponding gauge boson and the $\Delta X$
represent the BSM contributions.
\smallskip

At order $1/\Lambda^4$, the effective couplings
$\Delta \overline g_{i(W)}$ and $\Delta g^{\prime}_{i(W)}$ appearing
in \eqref{eq:dgZ} and \eqref{eq:dgW} depend upon the Wilson
combinations defined in Eqs.~\eqref{eq:cbwb}--\eqref{eq:c3whb} as
\begin{eqnarray}
  \Delta \overline{g}_1= &&
  - \frac{1}{4} 
  \left[ 2 \overline{\Delta}_{4F} 
    + \overline{c}_{\Phi,1} \right] \frac{\vh^2}{\Lambda^2}
 \label{eq:dg1b} \;, \\
\Delta \overline{g}_2=
&&-
\frac{\sh_2}{8\ch_2} \Big[
  \sh_2 \left(2 \overline{\Delta}_{4F} 
  + \overline{c}_{\Phi,1}\right)
  +4
  \overline{c}_{BW} \Big] \frac{\vh^2}{\Lambda^2} \;,
\label{eq:dg2b}\\
 \Delta \overline{g}_W=&&
 -\frac{1}{4\ch_2}  \left [
      2\sh_2 \overline{c}_{BW}
        +2 \ch^2 \overline{\Delta}_{4F}
        +\ch^2 \overline{c}_{\Phi,1} \right]\frac{\vh^2}{\Lambda^2}
 -\frac{1}{2}   \overline{c}^{(3)}_{W^2H^4} \frac{\vh^4}{\Lambda^4} \;,
\label{eq:dgwb}
\end{eqnarray}
and
\begin{eqnarray}
  \Delta g^\prime_1&&=\frac{\eh^4}{8\ch^4\sh^4}
\left( \sh^2 c^{(2)}_{\psi^4D^2} + \ch^2 c^{(3)}_{\psi^4D^2}\right)
\frac{\vh^4}{\Lambda^4}
\;,
\label{eq:dg1p}
\\
\Delta g^\prime_2&&=
\frac{\eh^4}{{ 8\ch^2\sh^2}}
\left( c^{(3)}_{\psi^4D^2} -c^{(2)}_{\psi^4D^2}\right)\frac{\vh^4}{\Lambda^4}\;,
\label{eq:dg2p}\\
\Delta g^\prime_W &&=
\frac{\eh^4}{8\sh^4}
c^{(3)}_{\psi^4D^2} \frac{\vh^4}{\Lambda^4} \;.
\label{eq:dgwp}
\end{eqnarray}
with $\hat{c}_{n}=\cos (n\hat{\theta})$ and
$\hat{s}_{n}=\sin (n\hat{\theta})$.  The coefficients
$\Delta g^{\square}_{i(W)}$ contain the additional contributions which
are quadratic in the dimension-six Wilson coefficients.  Their
explicit expressions as a function of the effective coefficients can
be found in Ref.~\cite{Corbett:2025oqk}.\smallskip

In addition to the corrections to the couplings of $Z$ and $W$ bosons
to fermion pairs, there are seven contact contributions to
four-fermion amplitudes:
\begin{itemize}
\item two at dimension six: $c_{2JW}$  and $c_{2JB}$ \;,
\item five at dimension eight : $c^{(2)}_{\psi^4 D^2}$ ,
  $c^{(3)}_{\psi^4 D^2}$, $c^{(4)}_{\psi^4 H^2}$,
  $c^{(5)}_{\psi^4 H^2}$, and $c^{(7)}_{\psi^4 H^2}$.
\end{itemize}
However, in DY, the contribution of $c_{2JB}$ and
$c^{(4)}_{\psi^4 H^2}$ ($c_{2JW}$ and $c^{(5)}_{\psi^4 H^2}$) always
enter together in the same combination. Furthermore, as before, the
dimension-six Wilson coefficients $c_{BB}$ and $c_{WW}$ induce a shift
on the gauge coupling constants which results in an associated shift
on the four-fermion operator coefficients.  Hence, we find that the
four-fermion contact amplitudes depend upon the two combinations:
\begin{eqnarray}
  \overline{c}_{2JW}
  &\equiv& c_{2JW}(1{-2\frac{\vh^2}{\Lambda^2} c_{WW}})
  + { c^{(5)}_{\psi^4 H^2} \frac{\vh^2}{2 \Lambda^2}}
  =-\frac{2\sh^2}{\eh^2}\overline{\Delta}_{4F} \;,
\label{eq:c2jwbap}
  \\
  \overline{c}_{2JB}&\equiv& c_{2JB}(1-2\frac{\vh^2}{\Lambda^2}c_{BB})
  + { c^{(4)}_{\psi^4 H^2}\frac{\vh^2}{2\Lambda^2}} \;.
\label{eq:c2jbbap}  
\end{eqnarray}
In addition $Q^{(7)}_{\psi^4 H^2}$ contributes to a different
combination of the momentum independent four-fermion contact
amplitudes.  Moreover, $Q^{(2)}_{\psi^4 D^2}$ and
$Q^{(3)}_{\psi^4 D^2}$ generate distinctive momentum-dependent
four-fermion vertices.
\smallskip

\section{Full matching}
\label{app:fullmatching}

The full matching results at dimension eight include many operators
not relevant to our analysis. We collect the full dimension-eight
Lagrangians here. In what follows we have \textit{suppressed the mass
  scaling} of the operators to improve the presentation. As an example
the following term in the Lagrangian for the mirror-$U(1)_Y$ model,
\begin{equation}
\mathcal L_8\subset  -\tfrac{1}{2}\beta^{2}\,Q^{(2)}_{\psi^{4}D^{2}}\, ,
\end{equation}
should be understood to scale as:
\begin{equation}
-\tfrac{1}{2}\beta^{2}\,Q^{(2)}_{\psi^{4}D^{2}}\to -\tfrac{1}{2M_1^2}\beta^{2}\,Q^{(2)}_{\psi^{4}D^{2}}\, .
\end{equation}

The full dimension-eight Lagrangian for the mirror-$U(1)_Y$ model is given by:
\begin{align}
  \mathcal{L}_{8} =& 
-\tfrac{1}{8}v^{2}g^{2}g^{\prime \,2}\beta^{2}(1+\beta^{2})\lambda_h\,Q_{\Phi,6}
+\tfrac{1}{4}g^{2}g^{\prime \,2}\beta^{2}(1+\beta^{2})\lambda_h\,Q_{H^{8}}\\
&-\tfrac{1}{2}g^{2}g^{\prime \,2}\beta^{2}(1+\beta^{2})\,Q^{(1)}_{H^{6}}
-\tfrac{1}{8}g^{2}g^{\prime \, 2}\beta^{2}(1+\beta^{2})\,Q^{(2)}_{H^{6}}
-\tfrac{1}{2}g^{\prime \,2}\beta^{2}(1+\beta^{2})\,Q^{(1)}_{H^{4}}    
+\tfrac{1}{2}g^{\prime \,2}\beta^{2}(1+\beta^{2})\,Q^{(2)}_{H^{4}}\nonumber\\
&+\tfrac{1}{16}g^{2}g^{\prime \,2}\beta^{2}(1+\beta^{2})\,Q^{(1)}_{W^{2}H^{4}}
-\tfrac{1}{16}g^{\prime\,4}\beta^{2}(1+\beta^{2})\,Q^{(1)}_{B^{2}H^{4}}
+\tfrac{i}{2}gg^{\prime\, 2}\beta^{2}(1+\beta^{2})\,Q^{(1)}_{WH^{4}D^{2}}
-\tfrac{i}{2}g^{\prime\,3}\beta^{2}(1+\beta^{2})\,Q^{(1)}_{BH^{4}D^{2}}\nonumber\\
&+\tfrac{1}{16}g^{2}g^{\prime \,2}\beta^{2}(1+\beta^{2})\,(Q^{(1)}_{\psi^{2}H^{5}}+Q^{\dagger(1)}_{\psi^{2}H^{5}})
-\tfrac{1}{8}gg^{\prime \, 2}\beta^{2}(1+\beta^{2})\,Q^{(2)}_{\psi^{2}H^{4}D}
+\tfrac{1}{4}g^\prime\beta^{2}\,Q^{(1)}_{\psi^{2}H^{2}D^{3}}\nonumber\\
&-\tfrac{3}{8}g^{\prime \, 2}\beta^{2}\,Q^{(1)}_{\psi^{4}H^{2}}  
+\tfrac{3}{8}g^{\prime \, 2}\beta^{2}\,Q^{(2)}_{\psi^{4}H^{2}}
+\tfrac{3}{8}g^{\prime \,2}\beta^{2}\,Q^{\dagger(2)}_{\psi^{4}H^{2}}
-\tfrac{3}{4}g^{\prime \, 2}\beta^{2}\,Q^{(3)}_{\psi^{4}H^{2}}
+\tfrac{1}{4}g^{\prime \, 2}\beta^{4}\,Q^{(4)}_{\psi^{4}H^{2}}
  -\tfrac{1}{2}\beta^{2}\,Q^{(2)}_{\psi^{4}D^{2}}\;.\nonumber
\label{eq:L8U1}
\end{align}
In the first line we also include a dimension-six operator,
$Q_{\Phi,6}=(H^\dagger H)^3$, which results from rewriting the purely
bosonic dimension-eight operators in terms of fermionic operators via
the EOM following \cite{Corbett:2024yoy}, in particular see the top of
Appendix A. The full set of operators appearing in this, and the
dimension-eight Lagrangian for the mirror-$SU(2)_L$ model can be found
in Tab.~\ref{fig:fulld8tab}.\smallskip

For the mirror-$SU(2)_L$ model we find:
\begin{align}
\mathcal{L}_8 =
&-\frac{1}{4}\, g^2 g^{\prime\,2}\, \beta^2(1+\beta^2)\, v^2 \lambda_h\, Q_{\Phi,6}
+\frac{1}{16}\, g^2\, \beta^2\!\left(-3 g^2 \beta^2+ 8 g^{\prime\,2}(1+\beta^2)\right)\! \lambda_h\, Q_{H^8}\\
&+\frac{1}{16}\!\left(9 g^4 \beta^4 - 12 g^2 g^{\prime\,2} \beta^2(1+\beta^2)\right)\! Q^{(1)}_{H^6}
-\frac{1}{2}\, g^2 g^{\prime\,2}\, \beta^2(1+\beta^2)\, Q^{(2)}_{H^6}\nonumber\\
&+\frac{1}{2}\, g^2\, \beta^2(1+\beta^2)\, Q^{(1)}_{H^4}
+\frac{1}{2}\, g^2\, \beta^2(1+\beta^2)\, Q^{(2)}_{H^4}
- g^2\, \beta^2(1+\beta^2)\, Q^{(3)}_{H^4}\nonumber\\
&+\frac{3}{16}\, g^2 g^{\prime\,2}\, \beta^2(1+\beta^2)\, Q^{(1)}_{B^2 H^4}
-\frac{1}{16}\, g^4\, \beta^2(1+3\beta^2)\, Q^{(1)}_{W^2 H^4}
+\frac{1}{8}\, g^3 g^\prime\, \beta^2\, Q^{(1)}_{W B H^4}\nonumber\\
&+\frac{3 i}{2}\, g^2 g^\prime\, \beta^2(1+\beta^2)\, Q^{(1)}_{B H^4 D^2}
-\frac{i}{2}\, g^3\, \beta^2(1+3\beta^2)\, Q^{(1)}_{W H^4 D^2}\nonumber\\
&
+\frac{1}{8}\, g^2 g^{\prime\,2}\, \beta^2(1+\beta^2)\, \left(Q^{(1)}_{\psi^2 H^5}+ Q^{\dagger(1)}_{\psi^2 H^5}\right)
-\frac{1}{2}\, g^2 g^\prime\, \beta^2(1+\beta^2)\, Q^{(1)}_{\psi^2 H^4 D}
+\frac{1}{4}\, g^3\, \beta^4\, Q^{(2)}_{\psi^2 H^4 D}\nonumber\\
&+\frac{1}{4}\, g^3\, \beta^4\, Q^{(4)}_{\psi^2 H^4 D}
+\frac{1}{4}\, g\, \beta^2\, Q^{(2)}_{\psi^2 H^2 D^3}
+ g^2\, \beta^2\, Q^{(11)}_{\psi^2 W H^2 D}\nonumber\\
&
-\frac{9}{8}\, g^2\, \beta^2\, Q^{(1)}_{\psi^4 H^2}
+\frac{3}{8}\, g^2\, \beta^2\, \left(Q^{(2)}_{\psi^4 H^2}+Q^{\dagger(2)}_{\psi^4 H^2}\right)
+\frac{3}{4}\, g^2\, \beta^2\, Q^{(3)}_{\psi^4 H^2}
+\frac{1}{4}\, g^2\, \beta^4\, Q^{(5)}_{\psi^4 H^2}
-\frac{1}{2}\, \beta^2\, Q^{(3)}_{\psi^4 D^2}
-\frac{1}{2}\, g\, \beta^2\, Q^{(1)}_{\psi^4 W} \;.\nonumber
\end{align}
As in the case of the mirror-$U(1)_Y$ model the dimension-six operator
$Q_{\Phi,6}$ is generated in changing from the purely bosonic to
fermionic basis. The full set of operators appearing in this
Lagrangian can be found in Tab.~\ref{fig:fulld8tab}.\smallskip

\begin{table}
\begin{tabular}{| rcl | rcl|}
\hline
\multicolumn{3}{|c|}{Bosonic Operators}&\multicolumn{3}{|c|}{Fermionic Operators}\\
\hline
$Q_{\Phi,6}$&=&$(H^\dagger H)^3$
&$Q_{\psi^2H^5}^{(1)}$&=&$(H^\dagger H)^2 (J_H H)$\\
$Q_{H^8}$&=&$(H^\dagger H)^4$
&$Q_{\psi^2H^4D}^{(1)}$&=&$i\,J_B^\mu (H^\dagger\overleftrightarrow{D}_{\mu} H) (H^\dagger H)$\\
$Q_{H^6}^{(1)}$&=&$(H^\dagger H)^2(D_\mu H)^\dagger (D^\mu H)$
&$Q_{\psi^2H^4D}^{(2)}$&=&$i\,J_W^{I\mu}\left[ (H^\dagger\overleftrightarrow{D}^I_{\mu} H) (H^\dagger H)+(H^\dagger\overleftrightarrow{D}_\mu H) (H^\dagger \tau^I H)\right]$\\
$Q_{H^6}^{(2)}$&=&$(H^\dagger H)(H^\dagger\tau^I H)(D_\mu H)^\dagger\tau^I (D^\mu H)$
&$Q_{\psi^2H^4D}^{(4)}$&=&$\epsilon^{IJK}\,J_W^{I\mu}(H^\dagger\tau^J_{\mu} H) D_\mu(H^\dagger \tau^K H)$\\
$Q_{H^4}^{(1)}$&=&$(D_\mu H)^\dagger (D_\nu H)(D^\nu H)^\dagger (D^\mu H)$
&$Q_{\psi^2H^2D^3}^{(1)}$&=&$i\,(D^\mu J_B^{\nu}+D^\nu J_B^{\mu}) (D_{(\mu}D_{\nu)} H^\dagger H-H^\dagger D_{(\mu}D_{\nu)} H)$\\
$Q_{H^4}^{(2)}$&=&$(D_\mu H)^\dagger (D_\nu H)(D^\mu H)^\dagger(D^\nu H)$
&$Q_{\psi^2H^2D^3}^{(2)}$&=&$i\,(D^\mu J_W^{I\nu}+D^\nu J_W^{I\mu}) (D_{(\mu}D_{\nu)}
  H^\dagger\tau^I H-H^\dagger\tau^I D_{(\mu}D_{\nu)}H)$\\
$Q_{H^4}^{(3)}$&=&$(D_\mu H)^\dagger (D^\mu H)(D^\nu H)^\dagger(D_\nu H)$
&$Q_{\psi^2WH^2D}^{(11)}$&=&$i\epsilon^{IJK}J_W^{I,\nu}(H^\dagger\overleftrightarrow D_\mu^J H)W_{\mu\nu}^K$\\
$Q_{B^2H^4}^{(1)}$&=&$(H^\dagger H)^2B_{\mu\nu}B^{\mu\nu}$
&$Q_{\psi^4H^2}^{(1)}$&=&$(H^\dagger H) J_H^j J^\dagger_{H,j}$\\
$Q_{W^2H^4}^{(1)}$&=&$(H^\dagger H)^2W_{\mu\nu}^IW^{I\mu\nu}$
&$Q_{\psi^4H^2}^{(2)}$&=&$(H_j J^j_H) (H_k J^k_H)$\\
$Q_{WBH^4}^{(1)}$&=&$(H^\dagger H)(H^\dagger\tau^I H)W_{\mu\nu}^I B^{\mu\nu}$
&$Q_{\psi^4H^2}^{(3)}$&=&$(H^{\dagger,k}J^\dagger_{H,k}) (J^j_H H_j)$\\
$Q_{BH^4D^2}^{(1)}$&=&$(H^\dagger H)(D^\mu H)^\dagger(D^\nu H)B_{\mu\nu}$
&$Q_{\psi^4H^2}^{(4)}$&=& $(H^\dagger H)J_B^{\mu}J_{B\mu}$   \\
$Q_{WH^4D^2}^{(1)}$&=&$(H^\dagger H)(D^\mu H)^\dagger\tau^I (D^\nu H)W_{\mu\nu}^I$
&$Q_{\psi^4H^2}^{(5)}$&=&$(H^\dagger H)J_W^{I\mu}J_{W\mu}^I$  \\
&&&$Q_{\psi^4D^2}^{(2)}$&=&$D^\alpha  J_B^\mu D_\alpha J_{B\mu}$\\
&&&$Q_{\psi^4D^2}^{(3)}$&=&$D^\alpha J_W^{I\mu} D_\alpha J_{W\mu}^I$\\
&&&$Q_{\psi^4W}^{(1)}$&=&$\epsilon^{IJK}W_{\mu\nu}^IJ_W^{J\mu}J_W^{K\nu}$\\
\hline
\end{tabular}
\caption{Full set of dimension-eight operators generated by the
  mirror-$U(1)_Y$ and mirror-$SU(2)_L$ models. We also include the
  dimension-six operator $Q_{\phi,6}$ as discussed in the text.}
\label{fig:fulld8tab}
\end{table}

\end{document}